\documentclass[pdftex,twocolumn,epjc3]{svjour3}          
\pdfoutput=1 
\RequirePackage[T1]{fontenc}
\usepackage{epsfig}
\usepackage{amsfonts}
\usepackage{slashed, enumitem}
\usepackage{enumerate}
\usepackage{float}
\usepackage[caption=false]{subfig}
\usepackage{multirow}
\usepackage{mathrsfs} 
\usepackage{amsmath}
\usepackage{amssymb}
\usepackage{slashed}
\usepackage{slashed}
\usepackage{psfrag}
\usepackage{float}

\usepackage[latin1]{inputenc}

\usepackage{amsfonts}
\usepackage{xcolor}
\usepackage{slashed}

\newcommand{\tr}{\text{tr}}

\journalname{Eur. Phys. J. C}
\begin{document}

  \title{\centering Inflation, (P)reheating and Neutrino Anomalies: Production of Sterile Neutrinos with Secret Interactions}   

\author{Arnab Paul\thanksref{e1,addr1}
	\and
	Anish Ghoshal\thanksref{e2,addr2,addr3} 
	\and
	Arindam Chatterjee\thanksref{e3,addr1} 
	\and
	Supratik Pal\thanksref{e4,addr1} 
}

\thankstext{e1}{e-mail: arnabpaul9292@gmail.com}
\thankstext{e2}{e-mail: anishghoshal1@gmail.com}
\thankstext{e3}{e-mail: arindam.chatterjee@gmail.com}
\thankstext{e4}{e-mail: supratik@isical.ac.in}

\institute{Physics and Applied Mathematics Unit, Indian Statistical Institute, 203 B.T. Road, Kolkata 700 108, India\label{addr1}
	\and
	Dipartimento di Matematica e Fisica, University Roma Tre, Via della Vasca Navale, 84, Rome-00146, Italy\label{addr2}
	\and
	Laboratori Nazionale di Frascati-INFN, C.P. 13, 100044, Frascati, Italy\label{addr3}
}


\maketitle
\begin{abstract}
 A number of experimental anomalies involving
neutrinos hint towards the existence of at least an extra (a very
light) sterile neutrino. However, such a species, appreciably
mixing with the active neutrinos, is disfavored by different
cosmological observations like Big
Bang Nucleosynthesis (BBN), Cosmic Microwave Background (CMB) and Large Scale
Structure (LSS). Recently, it was shown that the presence of
additional interactions in the
sterile neutrino sector via light bosonic mediators can make the
scenario cosmologically
viable by suppressing the production of the sterile neutrinos from
active neutrinos via
matter-like effect caused by the mediator. This mechanism works
assuming the initial
population of this sterile sector to be negligible with respect to
that of the Standard
Model (SM) particles, before the production from active neutrinos.
However, there is fair chance that
such bosonic mediators may couple to the inflaton and can be copiously produced
during (p)reheating epoch. Consequently, they may ruin this assumption
of initial small density of the
sterile sector. In this article we, starting from inflation,
investigate the production of such a sterile sector
during (p)reheating in a large field inflationary scenario and
identify the parameter
region that allows for a viable early Universe cosmology.
\end{abstract}

\newpage

\section{Introduction}
Several anomalies from different experiments measuring neutrino oscillations have 
hinted towards the existence of an additional sterile neutrino species. While 
\texttt{LSND} \cite{Athanassopoulos:1995iw,Aguilar:2001ty} and \texttt{MiniBooNE} \cite{Aguilar-Arevalo:2018gpe} reported an excess in $\bar{\nu}_{\mu}\rightarrow 
\bar{\nu}_e$ and the latter have also indicated an excess of $\nu_e$ in the $\nu_{\mu}$ 
beam.  Within a 3+1 framework, \texttt{MiniBooNE} result hints towards the existence of a 
sterile neutrino with eV mass at 4.8$\sigma$ significance, which raises to 6.1$\sigma$ 
when combined with the \texttt{LSND} data. Further, \texttt{Daya Bay} \cite{An:2016luf}, \texttt{NEOS} \cite{Ko:2016owz}, \texttt{DANSS} \cite{Alekseev:2016llm} and other reactor 
experiments \cite{Mention:2011rk,Mueller:2011nm,Huber:2011wv} probed the $\nu_e$ disappearance in the $\bar{\nu}_{e}\rightarrow \bar{\nu}_e$ channel, whereas  Gallium experiments \cite{Laveder:2007zz,Giunti:2006bj,Giunti:2010zu} like \texttt{GALLEX} \cite{Kaether:2010ag}, \texttt{SAGE} \cite{Abdurashitov:2009tn} have performed 
similar measurements in the $\nu_e\rightarrow\nu_e$ channel. The $\bar{\nu}_e$ disappearance 
data also hints in favour of sterile neutrinos at $3\sigma$ level.

However, there are significant tension among different neutrino experiments. In particular, 
observed excess in the experiments measuring $\nu_{\mu} (\bar{\nu}_{\mu}) \rightarrow 
\nu_{e} (\bar{\nu}_e)$ appearance (i.e. \texttt{LSND} and \texttt{MiniBooNE}) are in tension 
with strong constraints on $\nu_{\mu}$ disappearance, mostly from \texttt{MINOS} \cite{Adamson:2017uda} and 
\texttt{IceCUBE} \cite{TheIceCube:2016oqi}, while attempting to fit together using a 3+1 framework  \cite{Dentler:2018sju}.
Thus, the existence of a light ($\mathcal{O}(1)$ eV) sterile neutrino within 
a simple 3+1 framework, as a possible resolution to the $\nu_e$ appearance anomalies, 
remains debatable. \footnote{Recently a new idea involving sterile neutrinos altered dispersion relations was shown
to satisfy all the existing anomalies \cite{Doring:2018cob}.} However, such a light additional sterile neutrino, with mixing 
$\sin \theta \lesssim \mathcal{O}(0.1)$ with the active neutrino species, can be 
consistent with constraints from various terrestrial neutrino experiments. 

In the early Universe, production of a light sterile neutrino, if  exists, can be significant,  
thanks to its sizable mixing with the active neutrinos
 \cite{Barbieri:1990vx,Enqvist:1990ad,BARBIERI1990440,Kainulainen:1990ds,Dodelson:1993je,Merle:2015vzu}. Further, inflaton decays during re-heating, or any other heavy scalar 
particle can possibly decay into sterile neutrinos \cite{Shaposhnikov:2006xi,Merle:2013wta,Kusenko:2006rh}. Due to its sizable mixing with 
the active neutrinos, thermalization with the SM particles are also ensured. 
However, several cosmological constraints disfavor the viability of such a scenario. 
In particular, constraints from Big Bang Nucleosynthesis (BBN) \cite{Cooke:2013cba,pdg18,Cooke:2017cwo,BBN} restricts the effective 
number of relativistic degrees of freedom. This is because it would enhance the expansion 
rate at the onset of BBN ($\simeq 1$ MeV). Depending on its mass, such a species 
can contribute as either matter or radiation during matter-radiation equality epoch (for more explanation see \cite{nucos}). 
Additional non-relativistic neutrinos can also affect late-time expansion rate. Thus, 
it can affect the position of the acoustic peaks. Further, such light species can lead to 
slow down of Dark Matter clustering and thanks to their large free-streaming length \cite{silk}, can 
wash-out small-scale structure. Thus, Cosmic Microwave Background (CMB), together with 
Baryon Acoustic Oscillation (BAO) \cite{Alam:2016hwk,Zarrouk:2018vwy,boss} and Ly-$\alpha$ measurements \cite{Palanque-Delabrouille:2015pga,Yeche:2017upn} put forward significant 
constraint on the total neutrino mass $\Sigma m_{\nu}$ as well as on the number of relativistic 
degrees of freedom $N_{\rm eff}$ \cite{Mangano:2001iu}. Both of these constraints impact the viability of an 
additional light sterile neutrino species.

\texttt{Planck} \cite{Aghanim:2018eyx}, assuming three neutrinos with degenerate mass, Fermi-Dirac distribution and  zero chemical potential, constrains the properties of
neutrinos. If they become non-relativistic after recombination, they mainly affect CMB through the change of angular diameter distance, which is degenerate with $H_0$. 
So, the cleanest signal is through lensing power spectra that in turn affect the CMB power spectra. Since neutrino mass  suppresses lensing whereas CMB prefers higher
lensing, neutrino mass is strongly constrained by CMB lensing data. Neutrinos with large mass, that become non-relativistic around recombination, can produce distinctive 
features  in CMB (such as reducing the first peak height) and are thus ruled out.
\texttt{Planck} constrains $\Sigma m_{\nu}<0.12$ eV and $N_{\rm eff}=2.99_{-0.33}^{+0.34}$ for the 2018 dataset \cite{Vagnozzi:2017ovm,Aghanim:2018eyx} \texttt{Planck} TT + TE + EE + lowE + Lensing + BAO at 95$\%$ confidence level. It also constrains the effective mass of an extra sterile neutrino $m_{\nu, {\rm sterile}}^{\rm eff}<0.65$ eV with $N_{\rm eff}<3.29$ for the same dataset and same confidence level (though this value depends on chosen prior).
\footnote{In \texttt{Planck} 2015 results higher neutrino mass was allowed, though not favored. The effect of higher neutrino mass was nullified by larger primordial power spectrum amplitude $A_s$. This was allowed in \texttt{Planck} 2015 because of the degeneracy between optical depth $\tau$ and $A_s$, as a result of missing data points in low-multipole EE spectra, which could break the degeneracy.  \texttt{Planck} 2018 results constrain the $A_s$ to lower values, disfavoring higher $\Sigma m_{\nu}$.
}


While within the paradigm of standard cosmology all these constraints impact on the viability
of a light sterile neutrino with sizable active-sterile mixing, it has been shown that the CMB
constraints can be partly relaxed going beyond the standard cosmology modifying the primordial
power spectrum at small scales \cite{Canac:2016smv}, within the paradigm of modified gravity \cite{Chudaykin:2014oia}, within Beyond Standard Model (BSM) physics with time-varying Dark Energy component \cite{Giusarma:2011zq}, by light dark matter \cite{Ho:2012br}, from large lepton asymmetry \cite{Chu:2006ua}.
It has been also pointed out that additional interactions in the sterile neutrino sector
can also render such a scenario viable \cite{Babu:1991at,Enqvist:1992,Hannestad:2013ana,Archidiacono:2014nda,Archidiacono:2015oma,Archidiacono:2016kkh}. The presence of such interaction leads to the
suppression of the active-sterile mixing angle and thus delay the production. This significantly reduce the sterile neutrino abundance in the early Universe. In addition,
it provides a mechanism to cut-off the free-streaming length of the sterile neutrinos
at late time, and opens up an annihilation channel for the same. However, suppression of
this production alone does not suffice to constrain the energy density in the sterile neutrino
sector. In addition, one also needs to assume that post-inflationary production of sterile
neutrino and the light mediator, at least from the inflaton decay, remains small. During
the re-heating epoch, considering perturbative decay of the inflaton, this can be ensured
by simply assuming that the branching ratio of the inflaton into the sterile sector
particles remain insignificant compared to the Standard Model (SM) particles. However,
even this additional consideration does not serve the purpose when a bosonic mediator
is invoked. The reason is that post-inflationary particle production can be significant
during preheating \cite{Kofman:1994rk,Kofman:1997yn}. While light fermions, which couples to the inflaton, are not produced in
abundance during this epoch, the same does not hold for bosons.  A (light) boson, which
couples to the inflaton (via a quartic and/or tri-linear coupling, say) can be copiously
produced during this non-perturbative process, thanks to the large Bose enhancement.
Thus, while attempting to suppress the production of the sterile neutrino with secret
interactions, the possibility of producing the bosonic mediator during (p)re-heating,
and therefore, that of the sterile neutrinos can not be ignored.

In this article, we have considered a minimal renormalizable framework, consisting of an
inflaton, the Higgs boson, and the light mediator (interacting with the sterile neutrinos)
as only scalar particles to explore issues which are essential to make such a
sterile neutrino sector cosmologically viable, starting from inflation. Within this
framework all renormalizable terms are sketched out and their roles have been explored.
Generally, the inflaton couples to the light mediator which can give rise to large effective
mass during the inflationary epoch. Consequently, this prevents the light field to execute
jumps of order $\dfrac{H}{2 \pi}$, $H$ being the Hubble parameter during
inflation \cite{Starobinsky:1994bd}. Thus, it ensures any additional contribution to the energy density from the
light scalar remains negligible, which in turn, evades stringent constraint from
non-observation of iso-curvature perturbation by CMB missions \cite{Akrami:2018odb}.  However, the same
term can lead to the possibility of production during the preheating epoch. While the presence of a light scalar field with negligible
coupling to the inflaton have been considered in literature 
\cite{Starobinsky:1994bd}, and stringent constraint on the quartic
coupling of the light field have been put \cite{Enqvist:2014zqa}, aspects of the non-perturbative production,
especially with a small quartic self-coupling, during the preheating epoch
have not been considered in details. This may lead to serious issues which may destroy
inflationary cosmology altogether. In this article, we have explored the production of the scalar
mediator during (p)reheating and subsequent production of $\nu_s$, explicitly stressing
on regions of the parameter space where the production of the light mediator, and
consequently, the light sterile neutrino can be significant at the on-set of BBN, and also
the regions where such a sterile sector may be viable. We then discuss about some
benchmark parameter values elaborating the same.

The paper is organized in the following order. In section:\ref{model} the model has been
discussed. In the following section:\ref{preliminary_constraints} constraints on the relevant
model parameters from inflation, stability of the potential has been described. Further,
the constraints already present in the literature on such secret interactions has been sketched.  
Subsequently, in section:\ref{preheating} we discuss the production of the light pseudoscalar
during preheating and estimate the abundance of the sterile neutrinos in details. Finally, in
section:\ref{conclusion} we summarize our findings.


\section{Construction of the minimal framework}\label{model}
Before going into the model of inflation we remark that the interaction among the sterile neutrinos is same as in Ref. \cite{Archidiacono:2016kkh} - 
a pseudoscalar $\chi$ interacting with the sterile neutrino $\nu_s$ with the 
interaction
\begin{equation}
\mathcal{L} \sim g_s\chi\overline{\nu}_s\gamma_5 \nu_s. 
\end{equation}
The scalar part of the potential 
consisting of the fields inflaton $\phi$ (real scalar)
, $\chi$ and the SM Higgs $H$ is given by,
\begin{eqnarray}\label{simpleV}
	V &=& V_{\rm inf}+ \frac{1}{2}m_\chi^2\chi^2+\frac{\sigma_{\phi \chi}}{2} \phi \chi ^2 + \frac{\lambda_{\phi \chi}}{2} \phi^2 \chi ^2\nonumber\\ 
	&+&
	\frac{\sigma_{\phi H}}{2} \phi H^{\dagger}H+ \frac{\lambda_{\phi H}}{2} \phi^2 H^{\dagger}H + \frac{\lambda_{\chi H}}{2} \chi^2 H^{\dagger}H\nonumber\\ &+& \frac{\lambda_H}{4} (H^{\dagger}H)^2 + \frac{\lambda_{\chi}}{4}\chi^4 
\end{eqnarray}
where V$_{\rm inf}$ is the inflation potential. Reasons behind choosing an additional inflaton field $\phi$ instead of using $H$ or $\chi$ to drive inflation will be discussed in section:\ref{inflation}.
The final energy fraction transferred to $\chi$ or $H$ through $\lambda_{\phi \chi}\phi^2\chi^2$ or $\lambda_{\phi H}\phi^2 H^\dagger H$ interaction is weakly dependent on
the $\lambda_{\phi\chi}$ or $\lambda_{\phi H}$  couplings if the couplings are greater than a certain threshold value and energy is evenly distributed in $\phi$, $\chi$ and $H$ fields after preheating. 
Since back scattering $\phi$ particles to $\chi$ or $H$ is not much effective for energy transfer from the energy density left in inflaton to other 
sectors \cite{Dufaux:2006ee}, we consider trilinear term(s) for energy transfer typical of reheating. Inflaton-sterile neutrino coupling (of the form $\phi\nu_s\bar{\nu_s}$) does not help in the case we are 
concerned about, because it results in the total energy of the inflaton 
flowing into the sterile neutrino sector making the energy density in sterile sector and SM sector comparable, which ruins the $N_{\rm eff}$ bound at BBN. So we consider 
the trilinear terms $\frac{\sigma_{\phi \chi}}{2} \phi \chi ^2 $ and $
\frac{\sigma_{\phi H}}{2} \phi H^{\dagger}H$. We have neglected $\frac{\sigma_{\chi H}}{2} \chi H^{\dagger}H$ term, since it may give rise to mixing between $H$ and $\chi$ once Higgs get vev after EWPT. 
The inflaton decay rate arising due to a trilinear term $\frac{\sigma_{\phi \chi}}{2} \phi \chi ^2$ is given by,
\begin{eqnarray}
\Gamma_{\phi\rightarrow\chi\chi}=\frac{\sigma_{\phi\chi}^2}{16\pi m_{\phi}}\sqrt{1-\frac{4m_{\chi}^2}{m_{\phi}^2}}  
\end{eqnarray}	
Energy flow to any sector $i$ by decay of the inflaton depends on the branching ratio defined by, $\mathcal{B}_i=\frac{\Gamma_{i}}{\Sigma \Gamma_i}$.

\section{Cosmology and Light Sterile Neutrinos: Initial Constraints}\label{preliminary_constraints}
\subsection{Parameters of the scalar potential}

As observed in Section: \ref{model}, there are independent parameters in the minimal potential required for a viable cosmological scenario but in no way they can be of
arbitrary values. Rather, we expect them to be tightly constrained due to impositions by a successful inflationary paradigm as well as by phenomenological requirements 
of the current framework.

\subsubsection{Inflation, quantum corrections and threat to flatness of potential}\label{inflation}
 
We begin with exploring the requirements for a successful inflationary paradigm.
In this setup, primarily we have two scalar fields, namely, the SM Higgs and $\chi$. So, at first we argue why  we are using a separate inflaton field rather than using the scalar fields we already have, namely Higgs and $\chi$.



 Inflation with the Higgs field is a well studied subject \cite{Bezrukov:2007ep}. 
 It has been
shown that Higgs as inflaton requires a large non-minimal coupling of order $\xi\sim 50000$, which can result in so called unitarity violation \cite{Burgess:2009ea}. Moreover, if 
 $\lambda_{ H}$ becomes negative at high field values (which is certainly the case during inflationary energy scales), the non-minimal coupling 
is of no help to drive inflation.  However, some non-standard cases has been explored recently that enables the Higgs to be the inflaton. It has been found out that successful Higgs inflation can take place even if the SM vacuum is not absolutely 
stable \cite{Bezrukov:2014ipa}. It is also to be noted that if the action may be extended by an $R^2$ term on top of the Higgs non-minimal coupling to $R$; the Higgs field may drive inflation
\cite{Gorbunov:2018llf}. 
However, in this work we do not get into these scenarios and hence do not consider the Higgs to be the inflaton candidate. 

Using $\chi$ as inflaton is also problematic because large $\chi-H$ coupling is needed in
order to have enough energy flow to the Higgs sector (and thus to the SM sector) but
at the same time, this will induce a mass term to the $\chi$ field and consequently $\chi$ cannot
be light enough as required to evade $\Sigma m_\nu$ bound \cite{Archidiacono:2015oma}. Therefore we consider a separate inflaton field $\phi$.
 
 


In large field models of inflation, $\phi$ does not get a vacuum
expectation value at the end of inflation, and hence $\chi$ does not get effective mass from inflaton.
  For simplicity
 here we consider quadratic and quartic terms in the inflation potential.  However, driving inflation to produce the right amount of seed perturbation requires the
 inflaton potential to be very flat.
 In particular, in the large field inflationary models, e.g. $V(\phi)=(m_{\phi}^2/2)\phi^2$ or $V(\phi)=(\lambda_{\phi}/4)\phi^4$ one 
 requires $m_{\phi}\sim 10^{-6}$ M$_{Pl}$ or $\lambda_{\phi}\sim 10^{-14}$. Recent observations \cite{Akrami:2018odb} constrain the scalar power spectrum tilt $n_s\sim 0.9670\pm.0037$, as well 
 as the tensor perturbation via tensor to scalar ratio $r\leq0.065$ for the 2018 dataset \texttt{Planck}  TT,TE, EE + low E + Lensing + BK14 + BAO. Although $n_s-r$ bounds from  \texttt{Planck} observations does not satisfy the standard quadratic or quartic inflation 
 predictions, 
these observations can be satisfied if the inflaton $\phi$ is coupled non-minimally to gravity,
\begin{equation}
V_{J}=\frac{1}{2}\xi \mathcal{R} \phi ^2 +V_{\rm inf}
\end{equation}
Inflation constraints then can be easily satisfied with small values of $\xi \sim 10^{-3}$ to $1$ with $\lambda_{ \phi}\sim10^{-13}$ to $10^{-10}$ \cite{Kallosh:2013tua} for quartic inflation or  with $\xi\sim10^{-3}$, $m_\phi\sim10^{-6}$ M$_{pl}$ \cite{Linde:2011nh} for quadratic
inflaton. 
Therefore, to pursue this model of inflation, $\lambda_{ \phi}$ and $m_\phi$ are kept fixed to the values mentioned above and the rest of the parameters are varied
henceforth for our study.


Assuming the following inflaton potential $V_{inf} = \frac{m_{\phi}^2}{2}\phi^2+\frac{\lambda_{ \phi}}{4}\phi^4$ in Eq: \ref{simpleV},
where the symbols have their usual meanings, the Renormalization Group Equations (RGE) of the quartic interaction coefficients (assuming small non-minimal coupling $\xi$) for 1-loop look like \cite{Lebedev:2012zw,Ghosh:2017fmr},
\begin{eqnarray}
16\pi^2 {d \lambda_H \over dt}&=& 24 \lambda_H^2 -6 y_t^4 +{3\over 8} \Bigl( 
2 g^4 + (g^2 + g^{\prime 2})^2 \Bigr) \nonumber\\
&+& (-9 g^2 -3 g^{\prime 2}+12 y_t^2) \lambda_H + 2 \lambda_{\phi H}^2 + 2 \lambda_{\chi H}^2 \;, \nonumber\\
16\pi^2 {d \lambda_{\phi H} \over dt} &=& 8 \lambda_{\phi H}^2 + 12 \lambda_H \lambda_{\phi H}
+ 6 y_t^2  \lambda_{\phi H} + 6 \lambda_{\phi}  \lambda_{\phi H}\nonumber\\ &-&\frac{3}{2}(3 g^2 + g^{\prime 2}) \lambda_{\phi H}\;, \nonumber\\
16\pi^2 {d \lambda_{\phi} \over dt} &=& 8 \lambda_{\phi H}^2 + 2 \lambda_{\phi \chi}^2 + 18 \lambda_{\phi} ^2 \;,
\label{eq:rge}
\end{eqnarray}

From Eq: [\ref{eq:rge}], it may be concluded that if the value of $\lambda_{\phi}$ is  $\mathcal{O}(10^{-14})$ (for quartic inflation) at inflationary scales,
the terms in the RHS of the Eq: [\ref{eq:rge}] is needed to be less than or of same order of the 
initial value of $\lambda_{\phi}$ to keep $\lambda_{\phi}$ close to that order during the entire inflationary period. This means that the value of $\lambda_{\phi H}$ 
and $\lambda_{\phi \chi}$ to be of order $\lesssim \mathcal{O}(10^{-7})$ at inflationary
scales as $\lambda_{\phi H}$ does not evolve much with energy as clear from Eq: [\ref{eq:rge}]. These constraints on $\lambda_{\phi H}$ and $\lambda_{\phi \chi}$ can be weakened to $\leq \mathcal{O}(10^{-5})$ by increasing $\xi$ coupling to $\mathcal{O}(1)$ as a larger $\xi$ allows for higher values of $\lambda_{ \chi}$ for successful inflation. It should be noted that, in order to have quadratic inflation, i.e. to have the quadratic term dominating over the quartic term, $\lambda_{\phi}$ should be  $\lesssim\mathcal{O}(10^{-14})$. 
Similarly, the trilinear couplings $\sigma_{\phi H}$ and $\sigma_{\phi\chi}$ contributes to the running of $m_{\phi}$, which are negligible for small values of $\sigma_{\phi H},\sigma_{\phi\chi}\sim 10^{-8}-10^{-10}$ M$_{pl}$. Runnings of $\sigma_{\phi H}$ and $\sigma_{\phi\chi}$ are also insignificant for the small values of couplings discussed above \cite{Ema:2017ckf}. So, abiding by these constraints ensures us to get a successful inflation and preheating.

At high energy scales ($\sim10^{-5}$ M$_{\rm Pl} $), affected mostly by quantum corrections from top quark Yukawa coupling, the Higgs quartic coupling $\lambda_H$
becomes negative \cite{sher}. However, a positive value of $\lambda_H$ is required for a successful preheating phase, and will be shown in Section: \ref{preheating} that the energy flow to the Higgs sector during preheating explicitly depends on its 
value during preheating. Higgs stability is a well studied subject \cite{sher,EliasMiro:2011aa,Degrassi:2012ry,Buttazzo:2013uya} and ways to resolve the consequences during inflation are also well known \cite{Espinosa:2007qp,Enqvist:2014bua,Shkerin:2015exa,Lebedev:2012zw,Kearney:2015vba,Espinosa:2015qea,EliasMiro:2012ay}. As in this work we are only interested in a successful preheating dynamics, we shall assume that $\lambda_H$ stays positive during this era. 
This can be easily achieved with help of another
scalar coupled to Higgs, and as a result of this new coupling the new scalar thermalises
with SM sector. But we do not explicitly introduce the scalar in the discussion and
simply assume  $\lambda_H$ to be positive during preheating.

\subsubsection{Requirement of small mass for $\chi$ and $\nu_s$}
\label{vacuumstab}
If inflation is driven by quartic potential along with a trilinear term involving inflaton and another field (e.g. $\chi$) it gets vev at
the end of the inflation. This is problematic for model-buidling perspective as the vev of inflaton and $\chi$ 
result in mass terms for $\chi$ and $\nu_s$ respectively but we want the particles $\chi$ and $\nu_s$ to be of small masses $\mathcal{O}(eV)$. So, we would need
extreme fine tuning
in this case. On the other hand if the inflaton has a quadratic term then it is possible to have the minima of the potential at 0 field values. This is why we mainly consider quadratic inflation for our case. Nevertheless, even if we
choose to ignore the quartic term at some energy scale (i.e. set it to 0), it will become nonzero at other energy scales due to RGE running. 

In order to have stability of the potential we only need to check if the potential is bounded from below or not. But, in the present scenario we also need the 
minima of the potential to be at (0,0,0) for small mass of the $\chi$ particle without any fine tuning. As the potential Eq: \ref{simpleV} at (0,0,0) is 0, to have (0,0,0) as the 
minima, we need the potential to be non-negative and rewrite the potential as sum of positive terms:
\begin{eqnarray}
V &=& \frac{1}{2}\left(m_{\phi}\alpha\phi+\frac{\sigma_{\phi \chi}}{2m_{\phi}\alpha}\chi^2\right)^2+\frac{1}{4}\left(\lambda_{\chi}-\frac{\sigma_{\phi \chi}^2}{2\alpha^2 m_{\phi}^2}\right)\chi^4\nonumber\\&+&\frac{1}{2}\left(m_{\phi}\beta\phi+\frac{\sigma_{\phi H}}{2m_{\phi}\beta}H^{\dagger}H\right)^2\nonumber\\&+&\frac{1}{4}\left(\lambda_{H}-\frac{\sigma_{\phi H}^2}{2\beta^2 m_{\phi}^2}\right)(H^{\dagger}H)^2\nonumber\\&+& \frac{\lambda_{\phi \chi}}{2} \phi^2 \chi ^2+ \frac{\lambda_{\phi H}}{2} \phi^2 H^{\dagger}H +\lambda_{ \phi}\phi^4 +\frac{\lambda_{\chi H}}{2} \chi^2 H^{\dagger}H \nonumber\\
\end{eqnarray}
From this equation it is clear that the minima of the potential is at $(0,0,0)$ in field space (here we are talking about scenarios just after inflation and so we neglect
the EW vev of Higgs) under the conditions
$\lambda_{\chi}>\frac{\sigma_{\phi \chi}^2}{2\alpha^2m_{\phi}^2}$ and $\lambda_{H}>\frac{\sigma_{\phi H}^2}{2\beta^2m_{\phi}^2}$ given two arbitrary real constants $\alpha,~\beta$ with $\alpha^2+\beta^2=1$.  From these conditions, we can see that along with the quadratic inflation with $m_\phi\sim10^{-6}$, a scenario with small $m_\phi$ and small $\sigma_{\phi H}$ is equally possible with the inflation being a quartic one. 
In this work we choose the quadratic inflation case and try to understand the effects of the other parameters in the preheating dynamics in that case.
\subsubsection{Iso-curvature perturbations and stability of light fields during inflation}

Presence of light scalar fields during inflation can lead to iso-curvature perturbation 
\cite{Starobinsky:1994bd}.
The relaxation time scale for a quantum fluctuation to roll back down to its minima is $m^{-1}$ 
(during inflation, a field coupled to inflaton has mass depending on the expectation value of the inflaton $\phi_0$, if its bare mass is negligible), whereas
the time scale for the evolution of the universe is given by $H^{-1}$. So, if $m>H$, then the field is stable and the curvature	
perturbations due to that field may be neglected to be in agreement with constraint from
non-observation of iso-curvature perturbation by CMB missions \cite{Akrami:2018odb}. This condition translates to 
$\sigma_{\phi \chi}\langle\phi_0\rangle+\lambda_{\phi \chi}\langle\phi_0^2\rangle>m_{ \phi}^2\phi_0^2/M_{Pl}^2$ for quadratic inflation. Keeping $\sigma_{\phi \chi}/M_{Pl}$ or  $\lambda_{\phi \chi}$ substantially larger 
than 
$m_{\phi}^2/M_{Pl}^2$ can stabilize those fields during inflation. This is a condition we keep in mind while choosing these parameter values.

\subsection{Interaction parameters and bounds on $m_{\chi}-g_s$ plane}

For the sake of completeness of our discussion on different constraints on the parameters, 
let us  briefly summarize the existing constraints on the interaction parameters  from various physical requirements besides that of inflation and preheating. After the (p)reheating is over, the entire evolution of the spectrum(s) of the species is governed by the Boltzmann equation:
\begin{equation}
\mathcal{L}[f(E,t)]=C[f_i(E,t)]
\end{equation}
Here $\mathcal{L}$ is the Liouville operator and $C$ is the collision operator. We are interested in finding the spectrum of the sterile neutrinos through collision operators 
corresponding to $\chi\chi\longrightarrow\nu_s \nu_s$ and oscillation from active neutrinos. The Boltzman equation with the entire spectrum $f_i$ is difficult 
to solve even numerically, so it is assumed that the $\chi$ particles should follow a thermal distribution, i.e. $\chi$ particles are thermalized among themselves. For 
this assumption to be true just we need a parameter region of $\lambda_{\chi}$  estimated by the relation: 
\begin{equation}
 \Gamma > H;  \hspace{1cm}
 \Gamma \approx \langle\sigma v_{\rm mol}\rangle n,
\end{equation} 
where $\Gamma$ is the interaction rate, $H$ is the Hubble parameter, $\sigma$ is the interaction cross-section, $v_{\rm mol}$ is the Moller velocity of $\chi$ and 
n is the total number density. For our model, during radiation dominated epoch, for $\chi~\chi\rightarrow \chi~\chi$ scattering, $n_{\chi}=\frac{3}{4}\frac{\zeta(3)}{\pi^2}T_{\chi}^3 $ and $\sigma=\frac{4\pi}{64\pi^2 s}36 \lambda_{ \chi}^2\sim\frac{\lambda_{ \chi}^2}{T_{\chi}^2}$, gives $\Gamma\sim T_{\chi}\lambda_{ \chi}^2$;
whereas the Hubble parameter is given by $H=\sqrt{\frac{1}{3 M_{Pl}^2}\frac{\pi^2}{30}g_\star} T_{SM}^2$. As temperature of any relativistic species goes down at same 
rate $1/a$, even if the temperature of $\chi$ and SM are different, $\Gamma$ eventually becomes lower than H and gets thermal distribution. An estimate of the 
thermalisation temperature shows $\lambda_{ \chi}\sim 10^{-8}$ gives $T_{\rm Ther}\sim10^{-16}M_{Pl}$ (assuming $T_{SM}\sim T_{\chi}$, as even much lower energy density 
means temperature difference of order (density)$^{1/4}$), which is far before BBN.

The thermalization process within some sector starts much before the interaction rate (calculated from scattering of the particles) becomes comparable to the Hubble rate.
It is well known that at the start of the preheating epoch, modes with only some specific wave numbers gets excited exponentially governed by Mathieu equation. But, 
it has been observed from the \texttt{LATTICEEASY} simulation that, even if at the start of preheating stage, only some specific range of infrared momentum modes gets excited, as time 
progresses, the energy gets distributed to higher momentum modes. This observation can be interpreted as start of thermalisation process at the end of 
preheating \cite{Felder:2000hr}.
 
 The thermalisation of $\chi$ and $\nu_s$ is governed by the interaction 
 $\chi~\chi\rightarrow \nu_s~\nu_s$, having $\Gamma=\frac{3}{4}\frac{\zeta(3)}{\pi^2}T_{\chi}^3 \frac{g_s^4}{8\pi T_{\chi}^2}$. This means for $g_s\sim 10^{-4}$ the 
 thermalisation happens at 1 GeV \cite{Archidiacono:2014nda}.

\subsubsection*{Bounds on $m_{\chi}-g_s$ plane}\label{BBNbounds}
 
  \textbf{(i) From      BBN:} The standard way to parameterize the radiation energy density ($\rho_R$) is like \cite{Dodel},
 \begin{equation} 
  \rho_R=\rho_{\gamma}\left(1+\frac{7}{8}\left(\frac{4}{11}\right)^{4/3}N_{\rm eff}\right)\simeq\rho_{\gamma}(1+0.227N_{\rm eff})
  \label{BBNEqn}
 \end{equation}
 where $\rho_{\gamma}$ is the photon energy energy density, $N_{\rm eff}$ is the effective number of relativistic species. 
 A universe with only active neutrinos lead to $N_{\rm eff}=3.046$. Any extra radiation-like component (light sterile neutrinos), if present, will contribute to this
 $N_{\rm eff}$. 
 BBN observations constrain this value of $N_{\rm eff}$, we use a conservative bound of $N_{\rm eff}<3.5$ at $68\%$ confidence level \cite{nucos}.
%

To get an analytic solution for the evolution equation for $f_s$ (see for details  \cite{Barbieri:1990vx,Enqvist:1990ad,McKellar:1992ja,Sigl:1992fn}), the phase space distribution of $\nu_s$, in terms of the interplay of oscillations and 
collisions, we start with the equation,
\begin{eqnarray}
 \left( \frac{\partial}{\partial t}-HE \frac{\partial}{\partial E}\right) f_{s}(E,t)&=&\frac{1}{4}\sin^2(2\theta_M(E,t)\Gamma(E,t))\nonumber\\&\times&(f_{\alpha}(E,t)-f_s(E,t))
 \label{BE}
 \end{eqnarray}
where $f_{\alpha}$ is the distribution function of active neutrinos, $\Gamma$ is collision rate.
The presence of an effective potential $V_{\rm eff}$ (see Appendix for detailed calculation) leads to suppression of the neutrino production through MSW-like effect via the $\chi$ particle 
through the effective mixing angle $\theta_M$. 
The decay time  of $\chi$ into $\nu_s$ or $\nu$ particles in its rest frame is very small, but the time dilation does not allow them to decay 
before BBN \footnote{
	The decay of $\chi$ of mass $\sim 0.1$ eV occurs after the decay time scale $\gamma\Gamma^{-1}$ (where $\Gamma$  is the decay width of $\chi$ in its rest frame and $\gamma$ factor comes due to time dilation for a relativistic particle) and the age of the universe at some epoch is $\sim H^{-1}$. 
	Therefore, for $\chi$ to be still present during BBN, we require $\gamma\Gamma^{-1}>H_{BBN}^{-1}$, i.e. $\gamma>9.26\times10^{13} \frac{m_{\chi}}{eV} g_s^2$. Picking a conservative $g_s\sim 10^{-4}$ gives $\gamma>10^5$ which condition is easily satisfied for a $0.1$ eV particle during BBN ($\sim $ MeV). It is also to be noted that in the calculation we have ignored the mixing angle term, which will suppress the decay width further.}.

\begin{figure}
\epsfig{file=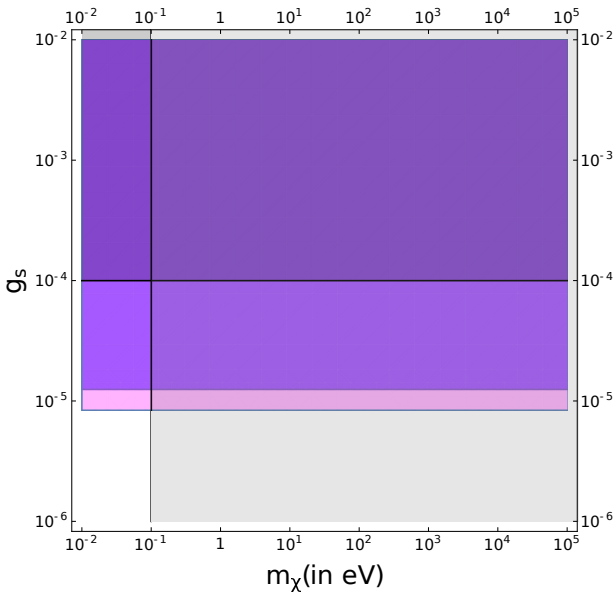,width= .95\linewidth}
 \caption{\it The blue and magenta regions correspond to the allowed regions in $m_{\chi}-g_s$ plane from $\triangle N_{\rm eff}$ (taking contribution form sterile neutrino only) constraints of BBN ($\triangle$ $N_{\rm eff}\lesssim0.5$) 
 for $\theta_0=0.1~and~0.05$; grey region above $g_s=10^{-4}$ is disfavored by 
 SNe energy loss bounds, whereas the grey region to the right of $m_{\chi}=0.1$ eV is disfavored from CMB, LSS bound on $\Sigma m_{\nu}$.}
\label{results1}
\end{figure}

From Eq: \ref{BE}, following  \cite{Jacques:2013xr}, the contribution of $\nu_s$ to $\Delta N^{\rm BBN}$ can be found to be
\begin{align}
&\Delta N^{\rm BBN}_{\rm eff, s} = \frac{f_s}{f_\alpha} \simeq  
1 - \exp \left[ \frac{-  2.06\times 10^3 }{\sqrt{g^\ast}}   \left(\frac{m_4 }{eV} \right)    \left( \sin ^2 \theta_{M}  \right) \right] .
\label{analytic1}
\end{align}
We expect the analytic expressions to be less accurate than numeric solutions mainly due to the discrepancy in $g^\ast$, which is kept fixed in the analytic
solution. The results achieved from analytics nearly resemble the full numerical results by solving quantum kinetic equations as in \cite{Archidiacono:2014nda}.
The Blue and magenta regions in Fig. \ref{results1} corresponds to the allowed region in $m_{\chi}-g_s$ plane from  BBN $N_{\rm eff}$ constraints for $\theta_0=0.1~and~0.05$.


 \vspace{.5cm}
 \textbf{(ii) From CMB \& LSS:} 
 The physical condition for the observed power spectrum in CMB and LSS is  not only to have the active neutrinos 
free-streaming, but also 
another sterile neutrino species (with $\mathcal{O}(eV)$ mass) not to be free-streaming. If this new species is of with similar number density as that of the active 
neutrinos, then there is much suppression in the power spectrum than 
that observed in CMB or LSS. So, to satisfy the CMB and LSS constraints of $\Sigma m_{\nu}$ (which basically quantifies the total mass of free-streaming species, assuming the same number 
density as that of active neutrinos), any massive species must annihilate or decay into lower mass particles to evade the mass constraints all together or they must interact among 
themselves or with some other species in order to cut off the free-streaming length scale. Ref. \cite{Archidiacono:2015oma} used the first recipe\footnote{ Ref. \cite{Chu:2015ipa} followed the second route, i.e., they used strong interaction strength for the 
	sterile neutrinos to self-interact via a secret mediator to cut off the free-streaming length. But, recent studies \cite{Chu:2018gxk} have shown that this scenario
	generates interactions between the active neutrinos too, through flavor mixing (note that the suppression in oscillation is lifted off for most of the parameter space for eV scale and below), 
	leading to a higher amplitude in power spectrum than the vanilla model itself. So, having a large interaction strength (with help of large coupling and/or small mediator mass) is 
	perilous for the cosmological observables, if one relies only on cutting off the free-streaming length.} to evade the mass 
bounds coming from CMB and LSS observations. They chose the mediator mass to be $\mathcal{O}(<0.1~eV)$ which meant that the interaction $\nu_s~\nu _s \rightarrow \chi~\chi$ goes only in the forward 
direction once the temperature of the Universe goes below $\mathcal{O}(1~eV)$, i.e. the backward interaction becomes kinematically inaccessible due to Hubble expansion dominating over
this process. Thus, for this suitable choice, the free-streaming length scale (as per CMB and LSS) need not be 
cut off since the sterile neutrinos annihilates into $\chi$ particles only, thereby not hurting the mass bound.
The gray region to the right of $m_{\chi}=0.1$ eV in Fig. \ref{results1} shows the excluded region by such arguments \cite{Archidiacono:2015oma}.

 \vspace{.5cm} 
 \textbf{(iii) From Supernova:}
 Constraints from SNe observations \cite{Farzan:2002wx,Heurtier:2016otg} also does not allow the couplings to be $g_s>10^{-4}$ \cite{Archidiacono:2016kkh}. So, this leaves us with a patch of parameter space at the 
 lower left corner of the BBN allowed region as depicted in Fig. \ref{results1}.

\section{Production of Sterile Sector from (P)reheating}\label{preheating}
Now, as the initial requirements are well understood, we discuss the production mechanism of different sectors. After the inflation is over, all the energy density is in the inflaton field, this energy density flows to other sectors by mechanisms like (p)reheating.
In this epoch, there are two stages of evolution:

\begin{itemize}
\item Initial stage that can be treated mostly analytically.
\item Back-reaction dominated stage for which one needs a detailed numerical treatment.
\end{itemize}
In what follows we shall analyse the two stages separately in order to get a complete picture.

\subsection{Initial stage of (P)reheating}
In this section we try to understand the mechanism of  energy flow from the inflaton to the other fields coupled to it by preheating. For simplicity we first 
consider the evolution of the inflaton (after inflationary period is over) neglecting the couplings to other fields. The evolution of the zero mode of inflaton is 
governed by,
\begin{eqnarray}
\ddot{\phi}+3H\dot{\phi}+V'(\phi)=0
\;, \nonumber\\
H^2=\frac{8\pi}{3M_p^2}\left(\frac{1}{2}\dot{\phi}^2+V(\phi)\right)\nonumber
\label{eq:inf0}
\end{eqnarray}
For quadratic potential it has sinusoidally oscillatory solution with decaying amplitude (due to Hubble friction term).

The dynamics of the Fourier modes of a field $\chi$ coupled to the inflaton in FRW universe is given by \cite{Kofman:1994rk},
\begin{equation}
\ddot{\chi}_k+3H\dot{\phi}_k+\left(\frac{k^2}{a^2}+\lambda_{\phi \chi}\Phi(t)^2\sin^2(m_{\phi}t)\right)\chi_k=0
\label{eq:fluc}
\end{equation}
This can be written (neglecting expansion) as the well known Mathieu equation, $$\chi_k''+\left(A(k)-2q\cos(2z)\right)\chi_k=0,$$where $A=\frac{k^2}{m_{\phi}^2 a^2}+2q$, $q=\frac{\lambda_{\phi \chi}\Phi(t)^2}{4m_{\phi}^2}$, $z=m_{\phi}t$ 
and prime denotes differentiation with respect to $z$.

Mathieu equation has well known unstable exponential solutions for instability regions of $A-q$ parameter space and hence for specific $k$ values. 
Modes corresponding to these $k$ values grow exponentially, which is interpreted as exponential particle production in those modes.

The statements up to now is only true for the initial stages of preheating. The growth of the fluctuations give rise to a mass term 
$\lambda_{\phi\chi}\langle\chi^2\rangle$ 
to the \textit{e.o.m} of the zero mode of the inflaton and as a result affects other modes through Eq: \ref{eq:fluc}. This phenomenon is known as the back-reaction 
effect \cite{Kofman:1997yn}; after the back-reaction effect starts, Eq: \ref{eq:fluc} does not hold true.
The back-reaction effect is usually estimated by the Hartree approximation, but still it does not take care of effects like re-scattering. So the only way to 
fully solve the\textit{ e.o.m} of the fluctuations throughout the preheating period is by lattice simulation. These simulations solve the classical field 
equations in lattice points numerically and give far accurate results than approximate analytic solutions.
\subsection{Numerical evolution for Back-reaction dominated stage}\label{numpre}


 To simulate preheating evolution numerically we use the publicly available code \texttt{LATTICEEASY} \cite{Felder:2000hq}.
 In this section we start from the 
simplest potential and discuss the results, motivating towards the final model. In each case we discuss the pros  and cons of the potential in hand. 
As we proceed we add new 
terms to the potential and clarify the implicit assumptions (as mentioned earlier) needed to reconcile the model with cosmological observations. It is to be noted that, we
neglect the effect of non-minimal coupling to gravity on the potential during the (p)reheating era, as a small coupling suffices to bring the inflationary scenario to the sweet spot of $n_s-r$ plane for a quadratic potential of inflaton. This is a logical assumption, as the potential remains unchanged near the minima of the 
inflaton for small non-minimal coupling, and the preheating is efficient when the inflaton oscillates about its minima.

	We start our discussion with the potential,
	\begin{eqnarray}
	V &=& \frac{m_{\phi}^2}{2}\phi^2+\frac{\lambda_{ \phi}}{4}\phi^4 + \frac{\lambda_{\phi \chi}}{2} \phi^2 \chi ^2 +
	\frac{\lambda_{\phi H}}{2} \phi^2 H^{\dagger}H \nonumber\\&+& \frac{\lambda_{\chi H}}{2} \chi^2 H^{\dagger}H
	+ \frac{\lambda_H}{4} (H^{\dagger}H)^2 + \frac{\lambda_{\chi}}{4}\chi^4 
	\end{eqnarray}
	We assume $\lambda_{\chi H}$ to be negligible, lest this parameter may thermalise the SM sector with that of the sterile. The energy density in the inflaton fluctuations, 
	Higgs and $\chi$ field at the end of preheating is of the same order if $\lambda_{ H}$ and $\lambda_{ \chi}$ are small with respect to $\lambda_{\phi H}$ and 
	$\lambda_{\phi\chi}$. 
	 We have observed from the simulation that, the energy-flow to a species is not only dependent on the coupling of the field with inflaton \footnote{If the coupling is lower than a certain 
	 threshold value, the preheating becomes inefficient (see Ref. \cite{Hardy:2017wkr} for details), 
	 whereas for values of that coupling over the threshold, the amount of energy flow is weakly dependent on 
	 the coupling}, but also on the self-quartic coupling of the field. A self-quartic coupling blocks the energy flow to that sector, as evident from
	 the plots (Fig:\ref{bm_2}). This salient feature is also in agreement with the studies in Refs.\cite{Hardy:2017wkr,Hyde:2015gwa}.
	 The reason for this phenomenon is the extra energy cost due to the potential term, which blocks the modes to grow. It can also
	 be interpreted from a different point of view: as the Fourier modes of $\chi$ grow, the quartic behaves like an effective mass term
	 $\frac{1}{2}\lambda_{\phi\chi}\langle\chi^2\rangle\chi^2$, making it difficult for the particle to be produced. This piece of information is very vital for
	 our work as we control the flow to $\chi$ sector by this self quartic term.

	If there is no trilinear coupling of the fields to inflaton, the inflaton cannot fully decay to other fields,
	thus contributing a significant amount to the total energy density of the Universe \cite{Dufaux:2006ee}, as mentioned in Sec: \ref{model}.
	Therefore, in order to direct the flow of the energy density stored 
	in inflaton to other species, we introduce trilinear couplings of inflaton to other scalars, as evident from the choice of potential in Eq: \ref{simpleV}.
	In the subsections below we will discuss the plausible scenarios case by case.

	\subsubsection{Trilinear interactions of inflaton with Higgs only}
	
\begin{figure*}
\epsfig{file=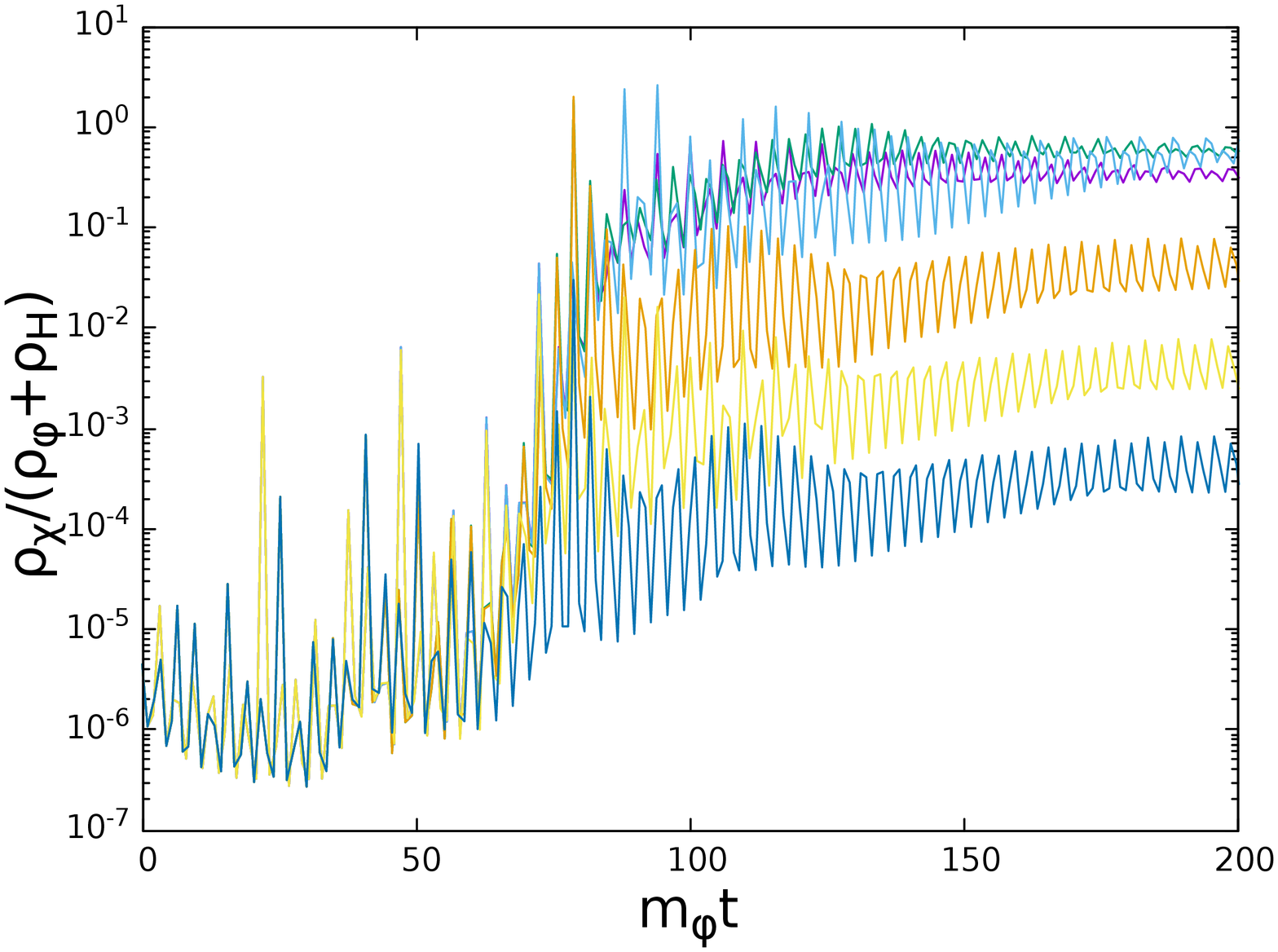,width= .47\linewidth}
\epsfig{file=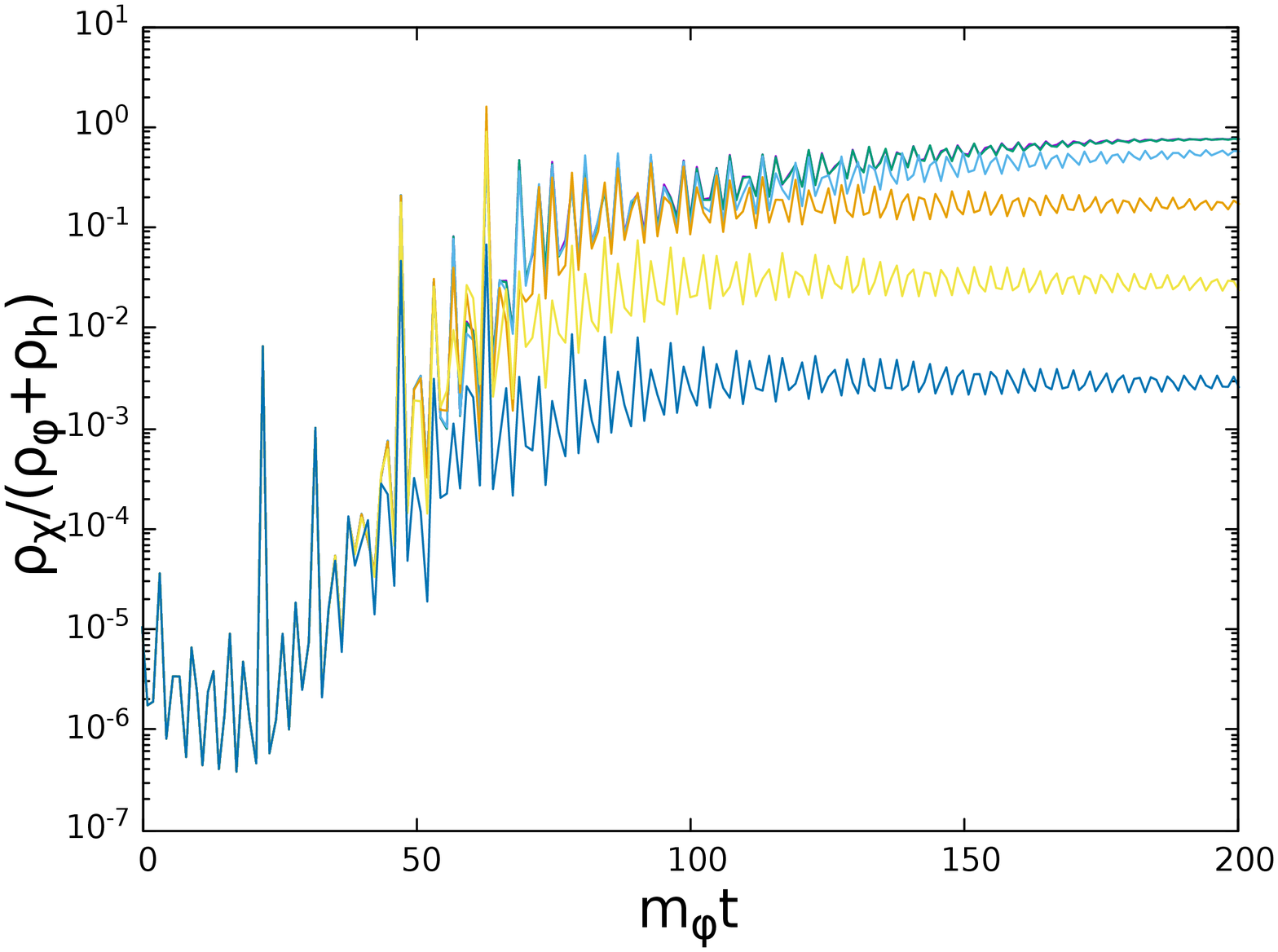,width= .47\linewidth}
	\caption{\it Plot of energy fraction ratios versus $m_{\phi}t$. 
	\textbf{Left:} $m_{\phi}=10^{-6}~M_{Pl},~\lambda_{\phi}=10^{-14},~\lambda_{\phi\chi}=10^{-7}$, $\lambda_{\phi H}=10^{-7},~\lambda_{H}=10^{-7}$, 
	$\sigma_{\phi H}=10^{-10}~M_{Pl}$. 
		\textbf{Right:} $m_{\phi}=10^{-6}~M_{Pl},~\lambda_{\phi}=10^{-14},~\lambda_{\phi\chi}=10^{-6}$, $\lambda_{\phi H}=10^{-6},~\lambda_{H}=10^{-6}$, 
		$\sigma_{\phi H}=10^{-10}~M_{Pl}$. $\lambda_{\chi}\sim 10^{-8} - 10^{-3}$ is varied (top to bottom). Higher value of $\lambda_{ \chi} $ leads to more 
		suppression of the $\chi$ energy fraction. }  
	
	\label{bm_2}
\end{figure*}

First we discuss a case where inflaton has trilinear interactions with Higgs only. In this case the total energy left in inflaton after preheating flows into SM sector. Even this case is not always compatible with cosmology if after preheating the order of energy densities in inflaton, Higgs and $\chi$ is same. To have a viable 
scenario we must find a way to block the production of $\chi$ during preheating - we do it by using the feature we observed in section: \ref{numpre} - a large self-quartic coupling 
in $\chi$ sector. 
This is one of our main results where we show 
that increasing the value of $\lambda_{ \chi}$ than $\lambda_{\phi \chi}$ increases 
the blocking of energy flow to $\chi$ sector, eventually resulting in lower energy density in this sector [Fig: \ref{bm_2}].
	
\begin{figure*}
\epsfig{file=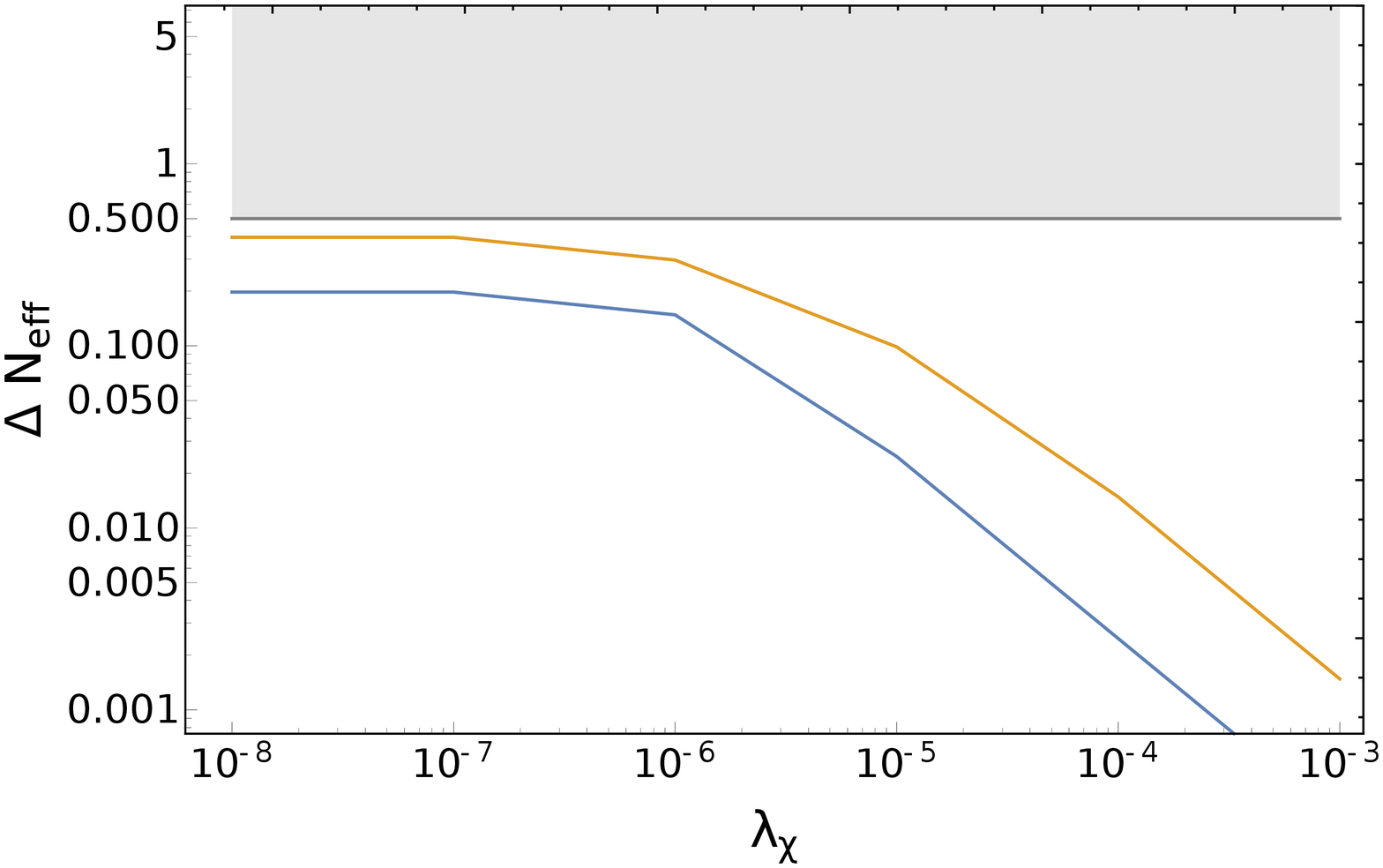,width= .32\linewidth}
\epsfig{file=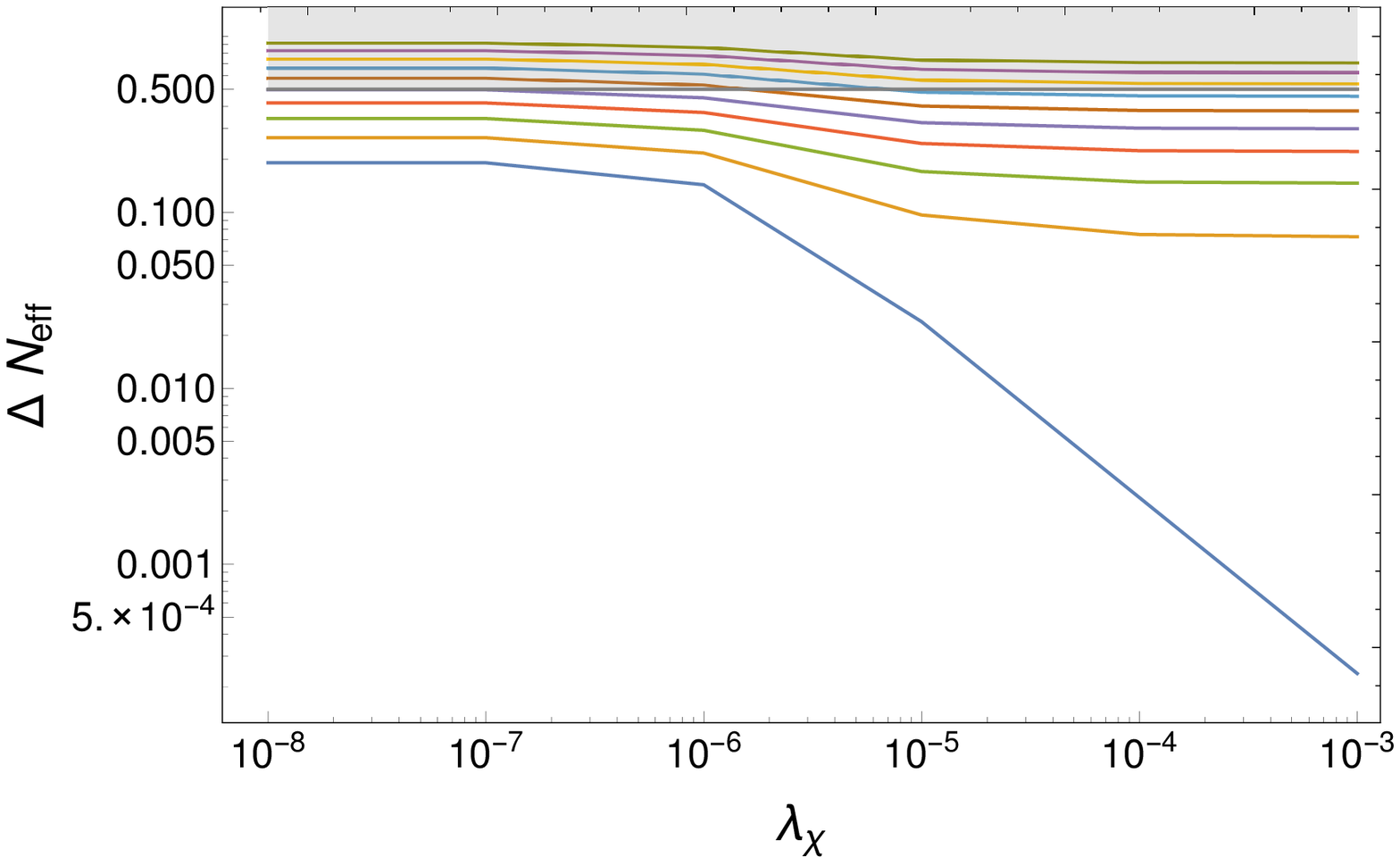,width= .32\linewidth}
\epsfig{file=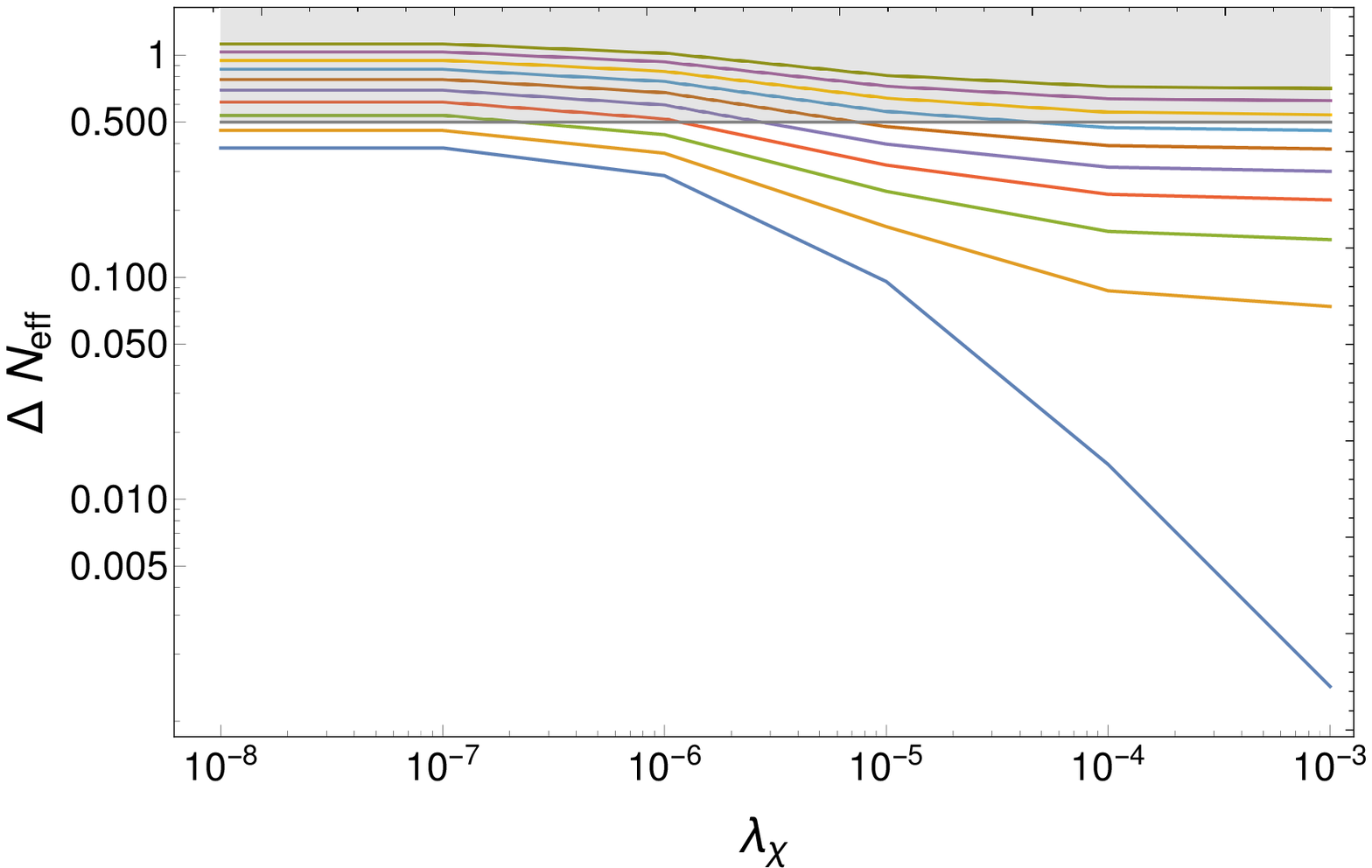,width= .32\linewidth}
		\caption{\it $\triangle N_{\rm eff}$ (taking contribution from both pseudoscalar and sterile neutrino)  in Y-axis versus $\lambda_{\chi}$ in X-axis, the left-most plot corresponds to the case when the inflaton 
		decays only into the Higgs. The parameter values, namely, are $m_{\phi}=10^{-6}~M_{Pl},~\lambda_{\phi}=10^{-14},
		~\lambda_{H}=10^{-7}$, $\sigma_{\phi H}=10^{-10}~M_{Pl},~\lambda_{\phi\chi}=\lambda_{\phi H}=10^{-7},10^{-6}$
		from bottom to top for the plot. Plots in the centre and right panels correspond to the cases $~\lambda_{\phi\chi}=\lambda_{\phi H}=10^{-7},10^{-6}$, when a fraction of the inflaton 
		(0 to 0.1 in steps of 0.01, from bottom to top)	decays into $\chi$ respectively. The grey region 
			($\triangle N_{\rm eff}>0.5$) is not allowed. }
			\label{Neff_lchinonrel}
	\end{figure*}

\begin{figure*}
\epsfig{file=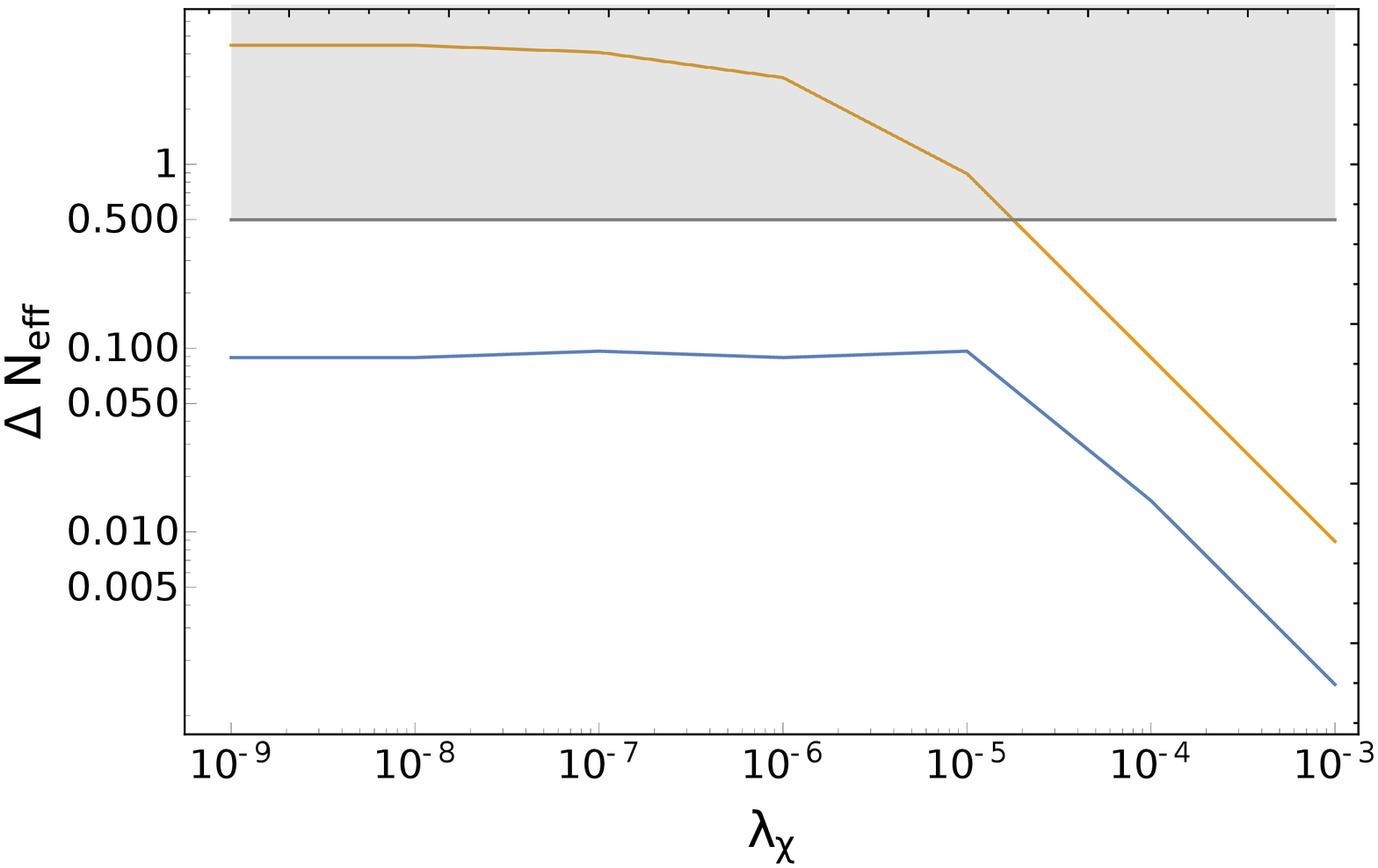,width= .32\linewidth}
\epsfig{file=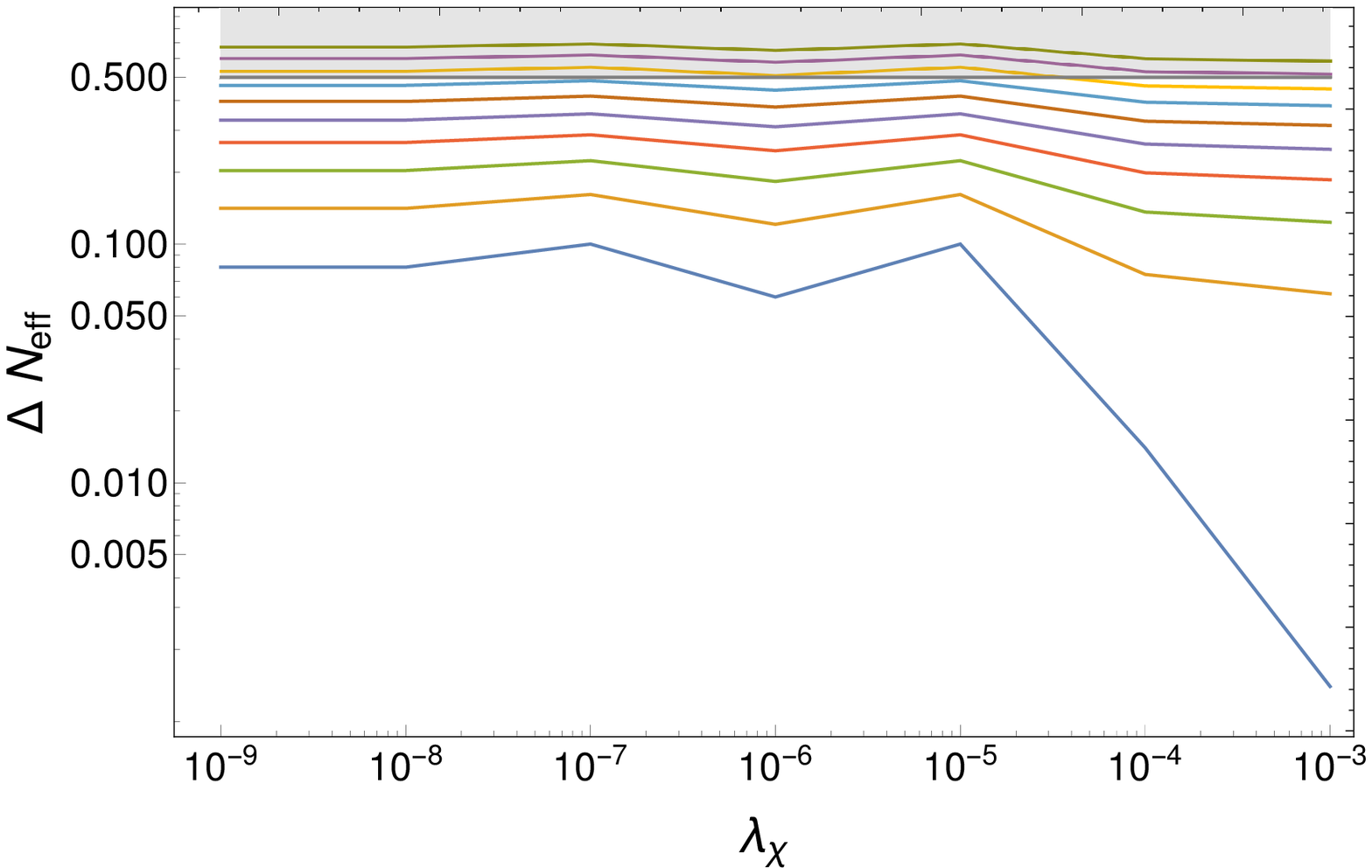,width= .32\linewidth}
\epsfig{file=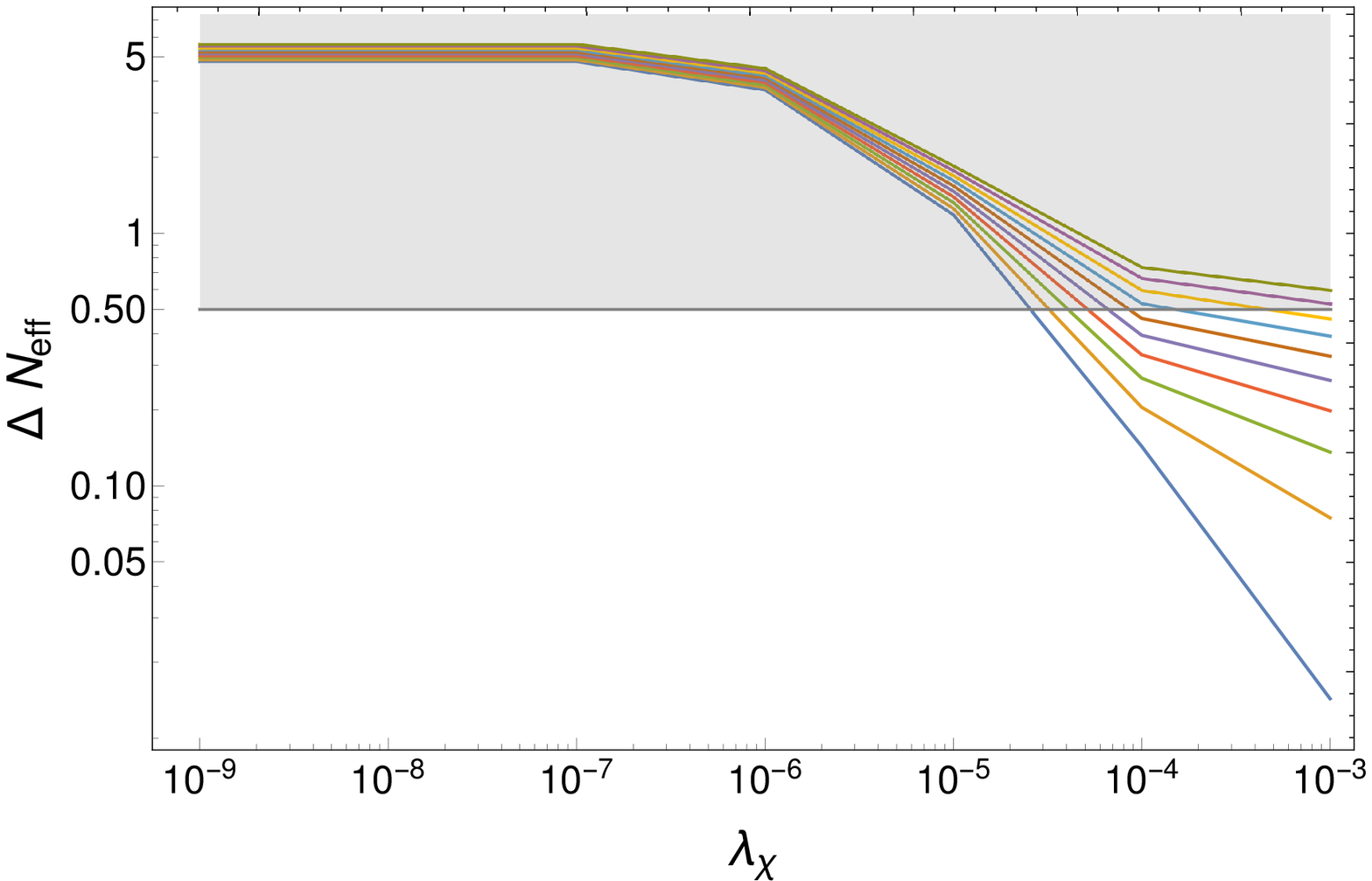,width= .32\linewidth}
\caption{\it $\triangle N_{\rm eff}$ (taking contribution from both pseudoscalar and sterile neutrino) in Y-axis versus $\lambda_{\chi}$ in X-axis, the left-most plot corresponds to the case when the inflaton 
	decays only into the Higgs. The parameter values, namely, are $m_{\phi}=10^{-6}~M_{Pl},~\lambda_{\phi}=10^{-14},~\lambda_{H}=10^{-4},~\sigma_{\phi H}=10^{-8}~M_{Pl},~\lambda_{\phi\chi}=\lambda_{\phi H}=10^{-7},10^{-6}$
	from bottom to top for the plot. Plots in the centre and right panels correspond to the cases $~\lambda_{\phi\chi}=\lambda_{\phi H}=10^{-7},10^{-6}$, when a fraction of the inflaton 
	(0 to 0.1 in steps of 0.01, from bottom to top)	decays into $\chi$ respectively. The grey region 
	($\triangle N_{\rm eff}>0.5$) is not allowed. }
	\label{Neff_lchi2}
\end{figure*}

The initial enhancement of energy density with time in $\chi$ sector upto $m_\phi t\sim 80$ resembles the part which can be described analytically by Mathieu equation as discussed before. For small $\lambda_{ \chi}$, the exponential increment in energy density stops when it becomes comparable to the energy density stored in the inflaton. This increment stops earlier for larger values of $\lambda_{ \chi}$.  Even if there is comparable energy density 
in the $\chi$, Higgs and inflaton sectors after preheating, the final relative energy density of the sectors depends on the state of inflaton (relativistic or non-relativistic) during its 
decay. 
Note that $T_{SM}$ decreases slower than $\frac{1}{a}$ due to entropy injection from heavy particles into lighter particles within SM sector.
	In our case, during preheating and subsequently from the decay of the inflaton the only SM particle produced is Higgs boson, thanks to its
	coupling to the inflaton at the tree-level. The energy density of Higgs then gets distributed into the other SM particles. Consequently 
	$g_{\star}$ increases to $106.7$ when reheat temperature is attained. Note that, even if initially the energy density of both SM and the
	sterile neutrino sectors were comparable, the temperature of the sterile sector can become higher than $T_{SM}$. At late time ($T_{SM}
	\simeq 100$ keV), $g_{\star}$ of SM decreases to  3.36. Thus, considering the entropy injection, at late time the ratio of energy density of the sterile
	sector to that of the SM is enhanced by $\frac{(g_{sterile}^{initial}/g_{sterile}^{final})^{1/3}}{(g_{SM}^{initial}/g_{SM}^{final})^{1/3}}=1.06$.
	In the above estimation we have assumed that
	$g_{sterile}^{initial},g_{SM}^{initial}=1$ since
	initially the only produced SM and sterile
	sector particles are Higgs and $\chi$ respectively\footnote{In this work, for computational simplicity, we have considered only one Higgs scalar degree of freedom (d.o.f), as is the case in the unitary gauge. However, we have checked for some points in parameter space, that considering four d.o.f of Higgs does not change our results significantly. This is because, in this case, the energy density   in each d.o.f of the Higgs doublet after preheating is lower than the scenario where only one scalar Higgs d.o.f is considered. The suppression in energy density in each Higgs d.o.f is caused by additional blocking from the cross-terms in the latter case. It should also be noted that, the fraction of Inflaton decaying into Higgs in the four d.o.f case results in a  Higgs sector with lower temperature. So, although there are four d.o.f, total energy density in the Higgs sector is not significantly different from the one d.o.f case, and hence our result does not change significantly (for example, for the parameter values $m_{\phi}=10^{-6}~M_{Pl},~\lambda_{\phi}=10^{-14},~\lambda_{H}=10^{-4},~\sigma_{\phi H}=10^{-8}~M_{Pl},~\lambda_{\phi\chi}=\lambda_{\phi H}=10^{-6},~\lambda_{ \chi}=10^{-9}$, $\triangle N_{\rm eff}$ changes from 4.8 to 3.5 for the four d.o.f case).}; also  $g_{sterile}^{final}=(\frac{7}{8}\times 2+1)=2.75$ and $g_{SM}^{final}=3.36$. Further, we
	have not taken into account the details of the thermalization post-preheating (it has been shown in \cite{Felder:2000hr} that the thermalization process starts at the end of preheating), and assumed comoving entropy conservation. 
. In Fig: \ref{Neff_lchinonrel} \& \ref{Neff_lchi2} (left panel) we plot $\triangle N_{\rm eff}$ for some benchmark values of the parameters (here $\triangle N_{\rm eff}$ corresponds to the whole sterile sector, i.e. pseudoscalar and sterile neutrino, where the sterile neutrino and pseudoscalar are thermalised, which is indeed the case for $g_s \sim 10^{-4}$. Note that in Fig: \ref{results1}, we considered only contributions from sterile neutrinos in $\triangle N_{\rm eff}$) and try to demonstrate the parameter values which satisfy the $N_{\rm eff}$ constraints at BBN. Note that, in this case
we have trilinear coupling of inflaton only to the Higgs, making sure that decay of inflaton does not populate $\chi$ sector. It is observed that even if a set of parameter values does not satisfy the $N_{\rm eff}$ constraints (Fig:\ref{Neff_lchi2} (left panel), top plot for small $\lambda_{ \chi}$), increasing the self-quartic coupling $\lambda_{ \chi}$ only can give a viable cosmological scenario again.
	
\begin{figure}[]
	\epsfig{file=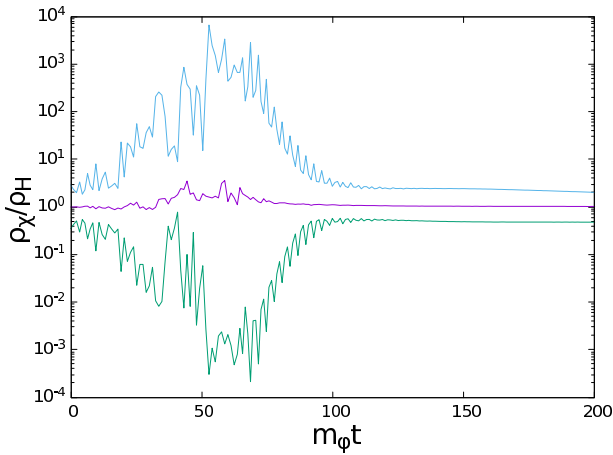,width= .95\linewidth}
	\caption{\it Plot of energy fraction ration of $\chi$ to that of Higgs versus $m_{\phi}t$, for fixed 
		$m_{\phi}=10^{-6}~M_{Pl},~\lambda_{\phi}=10^{-14},~\lambda_{\chi}=10^{-7}$, $\lambda_{H}=10^{-7}$, $\sigma_{\phi H}=10^{-10}~M_{Pl}$. The three plots
		from top to bottom correspond to ($\lambda_{\phi \chi},~\lambda_{\phi H}$) $\equiv$ ($10^{-6},~10^{-7}$), ($10^{-6},~10^{-6}$) and ($10^{-7},~10^{-6}$) respectively.
			}
	\label{lammix}
\end{figure}

We re-emphasise that the chosen values of the parameters are not arbitrary -  taking the
values of $\lambda_{\phi H}$ and $\lambda_{\phi \chi}$ too large can ruin the flatness of inflaton potential, 
whereas making them much smaller can result in inefficient preheating, $\lambda_{ \chi}$ and $\lambda_{ H}$ are the parameters we may vary to have a grip on the energy flow fractions in different sectors.
We have checked that keeping the 
$\lambda_{ H}$ and $\lambda_{ \chi}$ equal whilst $\lambda_{\phi H}$ and $\lambda_{\phi \chi}$ are unequal makes the energy flow to the Higgs and $\chi$ sector
as expected, i.e. the sector with higher $\lambda_{mix}$ ($\lambda_{\phi H}$ or $\lambda_{\phi \chi}$) will get larger share of energy density [Fig: \ref{lammix}].
After preheating, the energy left in the 
inflaton is transferred to Higgs sector through its perturbative decay due to the $\sigma_{\phi H}$ coupling. The inflationary model, if chosen to be quartic, 
the energy density stored in the inflaton evolves as radiation. Whereas for the quadratic case, the energy stored in the inflaton 
(the part behaving as a condensate) should behave as matter and this can dilute the energy densities produced in preheating. However it has been observed that, if a trilinear decay term 
is present, the equation of state behaves more like radiation than matter after the preheating stage \cite{Dufaux:2006ee}. The decay of the inflaton typically 
happens much after the preheating epoch. 
We assume that the inflaton becomes non-relativistic only after the temperature of the Universe becomes comparable to the mass of the inflaton. In our
analysis, we keep the values of $\sigma_{\phi H}$ of the order $\sim10^{-10}~M_{Pl}$ and $\sim10^{-8}~M_{Pl}$.
The choice of $\sigma_{\phi H} \sim 10^{-8}~M_{Pl}$ corresponds to the case when the T$_{R} \sim m_{\phi}$, i.e. there is no non-relativistic phase of inflaton before it decays. Whereas, if we decrease $\sigma_{\phi H}$, the 
decay of inflaton happens after it becomes non-relativistic, which means that
a matter dominated phase partly washes out the preheating contributions to the relativistic 
$\chi$ and Higgs and consequently, their final energy density, depending on the time span during which the inflaton is non-relativistic. Then the final energy 
density tends to solely depend on the 
branching fraction of inflaton, which is a trivial result. On the other hand, as shown in section: \ref{vacuumstab}, increasing $\sigma_{\phi H}$ and $\sigma_{\phi \chi}$
requires the increment of $\lambda_{ \chi}$ and $\lambda_{ H}$ values as well, which subsequently will lead to more blocking of corresponding energy density flows during preheating, 
thereby making the final energy	fractions dependent only on the branching ratios of the inflaton.
In the left most panels of Fig: \ref{Neff_lchinonrel} and \ref{Neff_lchi2} we show the $\triangle N_{\rm eff}$ corresponding to the sterile sector energy density for $\sigma_{\phi H}\sim 10^{-10}$ and $10^{-8}$. It is clear from Fig: \ref{Neff_lchinonrel} that even if $\chi$ is copiously produced from preheating, for $\sigma_{\phi H}\sim 10^{-10}$, the non-relativistic phase of inflaton before it decays to Higgs, can dilute the $\chi$ sector energy density produced from preheating and make the scenario cosmologically viable.	Whereas a larger $\sigma_{\phi H}\sim 10^{-8}$ (Fig: \ref{Neff_lchi2}), with no such non-relativistic phase, is highly constrained for $\lambda_{\phi \chi}=\lambda_{\phi H}\sim 10^{-6}$ at low $\lambda_{ \chi}$ values. The information we get from these plots is that smaller $\sigma_{\phi H}$ and higher $\lambda_{ \chi}$ is beneficial to get a cosmologically viable scenario. 


\subsubsection{Trilinear interactions of inflaton with $\chi $ only}
Inflaton with trilinear interactions with only $\chi$ is trivially not satisfied by the $N_{\rm eff}$ constraints as in this case the total inflaton energy after 
preheating flows into the $\chi$ sector.

	\subsubsection{Trilinear interactions with both Higgs and $\chi$}
\begin{figure}
	\epsfig{file=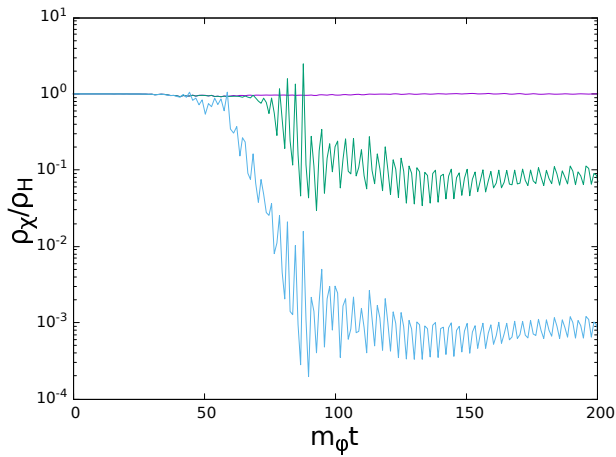,width= .95\linewidth}
	\caption{\it Energy fraction of $\chi$ with respect to that of the SM (basically the Higgs) versus $m_{\phi}t$ plot. Choosing the following set of values:
	($m_{\phi}=10^{-6}~M_{Pl},~\lambda_{\phi}=10^{-14},~\lambda_{\phi\chi}=10^{-7}$, $\lambda_{\phi H}=10^{-7},~\lambda_{H}=10^{-7}$, 
	$\sigma_{\phi \chi}=10^{-10}~M_{Pl}$, $\sigma_{\phi H}=10^{-10}~M_{Pl}$). The plots from the top to bottom correspond to $\lambda_{\chi}=10^{-7},10^{-5}$ and 
	$10^{-3}$ respectively.}
	\label{bm_tribothpl}
    \end{figure}

In this case inflaton has trilinear interactions with both Higgs and $\chi$. At first during preheating both Higgs and $\chi$ are produced, and energy density 
ratios depend on the self quartic couplings of those sectors. Then depending on the mass of the inflaton and the trilinear couplings, the inflaton decays into Higgs and 
$\chi$ (and also possibly $\nu_s$) according to their branching fractions. In Fig:\ref{bm_tribothpl} we show the fraction of energy density of $\chi$ and Higgs during 
preheating when inflaton has same trilinear coupling to both $\chi$ and Higgs. We have checked that for the parameter values we used, the final energy fractions after preheating do not differ by
much, depending on the presence of trilinear term in the $\chi$ sector. This is due to the fact that the production during preheating through the trilinear term is sub-dominant. But even if the fraction of $\chi$ energy is small after preheating, in the presence of a trilinear term, the inflaton decay to $\chi$ channel is open
and for the chosen $\sigma_{\phi \chi}=\sigma_{\phi H}=10^{-10}~M_{Pl}$ value, the branching ratio is same for both the fields (neglecting $m_H$ and $m_{\chi}$ to be negligible 
with respect to $m_{\phi}$). This case is not allowed by $N_{\rm eff}$ bounds of BBN. Varying the ratio of $\sigma_{\phi \chi}$ and $\sigma_{\phi H}$ changes the 
branching fraction and enables the Higgs and the $\chi$ sectors to be populated unequally during inflaton decay. In Fig: \ref{Neff_lchinonrel} (two panels from right), we show that even a small branching fraction of below 0.1 into the $\chi$ sector can be cosmologically prohibited. If we do not want an epoch when the inflaton behaves as a 
non-relativistic species (i.e., when the history of preheating dilutes away), then we would like to have 
large $\sigma_{\phi H}$ and $\sigma_{\phi \chi}$ ($\mathcal{O}(10^{-8})$ M$_{Pl}$). In Fig: \ref{Neff_lchi2} (two panels from right), we show how the decay channel of the inflaton into $\chi$ 
can change the $\triangle N_{\rm eff}$ values. However, in this case, the $H$ and $\chi$ sectors will thermalise with each other 
through a $HH\rightarrow\chi\chi$ scattering mediated through inflaton, resulting in a fully thermalised $\chi $ species with the SM, thereby trivially not respecting the $N_{\rm eff}$ bounds of BBN.

\subsection{Allowed benchmark parameter values and additional constraint on $m_{\chi}-g_s$ plane}
In our scenario, since we assume $m_{\chi}<2m_{\nu_s}$, 
 the primary production channel of the $\nu_s$ particles is from the $\chi$ via back-scattering
particles since the production from active neutrinos through oscillation is suppressed. The $\nu_s$ production from backscattering depends
on the $n \langle\sigma v_{\rm mol}\rangle $ of the interaction in comparison to the Hubble parameter. The thermally averaged cross-section for the process
$\chi\chi\rightarrow\nu_s\nu_s$ is given by (in the relativistic limit) \cite{Dolgov:1996fp}, 
\begin{equation}
\langle\sigma v_{\rm mol}\rangle=\frac{g_s^4}{8\pi T^2}
\end{equation}
So, the $\chi$ and $\nu_s$ sectors thermalise at $T\sim1$ GeV for $g_s\sim 10^{-4}$. The extra $N_{\rm eff}$ coming from the entropy of $\chi$ particles, and from the
$\nu_s$ particles, both of which are now partly or fully 
thermalised with each other depending on $g_s$, changes the bound in $m_{\chi}-g_s$ plane from BBN which was previously considered in the case for production of $\nu_s$ only from active 
neutrinos when inflation was not considered in the whole picture. We emphasise that a large enough initial abundance of $\chi$ particles (after (p)reheating) can make the total cosmological scenario not viable if $N_{\rm eff}$ bound is violated at the time of BBN, which is indeed the case for several regions of the parameter space, as clear from Fig: \ref{Neff_lchi2}. 
For the parameter region where the bound is satisfied, i.e.  $\triangle N_{\rm eff}\lesssim0.5$ , we get additional constraints in the $m_{\chi}-g_s$ plane reducing 
the allowed region in the parameter space of Fig. \ref{results1}.

\section{Conclusions and Outlook}\label{conclusion}
In this article we investigated for the possibility of having an extra sterile neutrino in the particle 
spectrum. It is well-known that it is possible to reconcile the 
cosmological observations with this extra species, if we introduce a pseudoscalar interacting with 
the sterile neutrinos. As a background field, the pseudoscalar particle
creates an effective thermal neutrino potential, which, due to its matter-like effect, suppresses the Dodelson Widrow-like 
production of sterile neutrinos. 
Such a BSM scalar, if present in the early universe, should also be produced inadvertently 
during preheating due to its coupling to the inflaton.
Moreover, Bose enhancement will make this production copious enough such that its primordial
abundance and that of sterile neutrinos cannot be assumed to be negligible with respect to SM particles for all scenarios. 
This assumption was the cornerstone of the scalar or vector interactions that alleviate the sterile neutrino
constraints from cosmology. Even though the production during reheating can be neglected by
considering small trilinear coupling as the inflaton decay is 
straight-forward and depends solely on the branching fraction, the production from preheating is non-trivial. Thus, it is important to consider 
the production right from preheating, which is a highly 
non-linear process, 
 this was studied  in exquisite details in this article.
We made use of analytical arguments and numerical calculations
using \texttt{LATTICEEASY} to find 
the regions of the parameters where this production will be significant. 
We observe that the pseudoscalar abundance from preheating, and hence the abundance of the whole sterile sector, can be high enough to violate the bounds of $N_{\rm eff}$ from BBN, 
if the self quartic coupling is not high enough and/or there is no period before decay of inflaton when it is non-relativistic. It is to be noted that even if these two conditions are satisfied, the inflaton can still decay to the pseudoscalar along with Higgs and the BBN $N_{\rm eff}$ bound can be at risk. We found benchmark values of the model parameters, for which the initial abundance
of the pseudoscalars and sterile neutrinos are favorable for a viable cosmological scenario, and discussed the impact of the various parameters on the final abundances of the
pseudoscalar and SM particles. It turns out that the for a small non-minimal coupling to gravity, an inflation mass (m$_{\phi}$ of $10^{12}$ GeV) compatible with 
recent $n_s - r$ observations of \texttt{Planck} 2018, and small self-quartic of inflaton ($\lambda_{\phi} \sim 10^{-14}$), we have the suitable parameter space of 
$\lambda_{\chi} \geq 2 \times 10^{-5}$ for mixing parameters $\lambda_{\phi \chi}$ and $\lambda_{\phi H}$ of the $\mathcal{O}(10^{-6})$. 
The trilinear term in the potential, $\sigma_{\phi H}$ was kept of order $\mathcal{O}(10^{-8})$ M$_{Pl}$ or smaller, chosen such that the inflaton decays when the temperature of 
universe is of the same order of the m$_{\phi}$ or latter. 
  To summarise, we restate  the following conclusions:
\begin{itemize}
	\item $\lambda_{\chi}$ needs to be kept large to suppress the preheating production of $\chi$.
	\item $\sigma_{\phi H}$ needs to be small in order to have a non-relativistic phase of inflaton before it decays and dilute the relic of $\chi$ from preheating. $\sigma_{\phi \chi}$ needs to be much smaller to prevent $\chi$ population during inflaton perturbative decay.
	\item  $\lambda_{\chi H}$ needs to be negligible to prevent SM from thermalizing with the BSM sector.
	\item $\lambda_{\phi H}$ and $\lambda_{\phi \chi}$ cannot be large, or else the flatness of inflaton potential will be ruined due to RG running.
\end{itemize}
 Thus the  sterile neutrino with the pseudoscalar ``secret-interaction'' model can be an viable possibility when all the early universe 
cosmology is considered on one hand, and, the existing neutrino anomaly is invoked on the other hand.

Future CMB missions like \texttt{LiteBIRD} \cite{lite}, \texttt{COrE} \cite{DiValentino:2016foa}, \texttt{PIXIE} \cite{pixie}, \texttt{CMB S4} \cite{Abazajian:2016yjj},
\texttt{CMB Bharat} \cite{cmbbharat} aim at constraining the inflationary observables and the other cosmological parameters further  and hence will be an important probe
for this model. Results from the neutrino experiments
 \texttt{MicroBooNE} \cite{Acciarri:2016smi} may decidedly prove the existence of the sterile neutrino. Sterile neutrinos with secret interactions have been 
proposed to be looked for in \texttt{IceCube} experiment \cite{Chauhan:2018dkd}. In CMB polarization observations of \texttt{BICEP} the sterile neutrinos may also be a relevant signature to
look for as per Ref. \cite{Choudhury:2018sbz}. 


\section{Acknowledgement}\label{Acknowledgement}

Authors gratefully acknowledge the use of the publicly available code, \texttt{LATTICEEASY} and also thank the computational facilities
of Indian Statistical Institute, Kolkata. A.P. thanks CSIR, India for financial support through Senior Research Fellowship (File no. 09/093 (0169)/2015 EMR-I).
A.C. acknowledges support from Department of Science and Technology, India, through INSPIRE faculty fellowship, (grant no: IFA
15 PH-130, DST/INSPIRE/04/2015/000110). A.G. is indebted to Davide Meloni for various support and encouragement and the hospitality of 
Indian Statistical Institute. Authors like to thank Gary Felder, Anupam Mazumdar, Alessandro Strumia and Subhendra Mohanty for discussions.

\section{Appendix}

\subsection{Suppression of $\nu_s$ production from active-sterile oscillation}
The evolution of spectrum of $\nu_s$, $f_{\nu_s}$(E,t) is governed by, assuming equilibrium distribution for active neutrinos,
\begin{eqnarray}
\left( \frac{\partial}{\partial t}-HE \frac{\partial}{\partial E}\right) f_{\nu_s}(E,t)&=&C_{\chi\chi\longrightarrow\nu_s \nu_s}\nonumber\\&+&\frac{1}{2}\sin^2(2\theta_M(E,t)\Gamma(E,t))\nonumber\\&\times&f_a(E,t)
\end{eqnarray} where
$\Gamma$ is \cite{Cline:1991zb,Notzold:1987ik}:
\begin{equation}
 \Gamma = 0.92 \times G_{\rm Fermi} ^2 E T^4 
\end{equation}
for active neutrino re-population and
\begin{equation}
 \Gamma = \frac{g^4 _{s}}{4\pi T ^2 _{s}} n_{\nu_s}
\end{equation}
for sterile neutrino redistribution through Fermi-Dirac distribution. The latter can be neglected since this is very small.
$C_{\chi\chi\longrightarrow\nu_s \nu_s}$ is the collision term corresponding to $\chi$ annihilation given by:
\begin{eqnarray}
 C_{\chi\chi\longrightarrow\nu_s \nu_s} (q) &=& \frac{1}{2 E_q} \int \int \int \frac{d ^3 q^{\prime}}{(2\pi)^3 2 E_{q^{\prime}}} \frac{d ^3 p}{(2\pi)^3 2 E_{p}} \frac{d ^3 p^{\prime}}{(2\pi)^3 2 E_{p^{\prime}}} \nonumber\\&\times&\sigma_{\chi\chi \rightarrow \nu_s \nu_s} (2\pi)^4 \delta_E \delta_p f^{eq}_q f^{eq}_{q^{\prime}}
 \label{B}
 \end{eqnarray}
Now there maybe two cases one where the particles are already thermalised, so that one may take $f^{eq}_q = e^{-\frac{q}{T}}$ and in another case where they are not 
in thermal equilibrium and need to be numerically evolved from preheating dynamics. The second term in \eqref{B} corresponds to an oscillation term, 
$\theta_M$ being the mixing angle which is suppressed by introduction of the hidden sector interaction through a effective potential $V_{\rm eff}$ as,
\begin{equation}
\sin^2(2\theta_M)=\frac{\sin^2(2\theta_0)}{\left( \cos(2\theta_0)+\frac{2E}{\delta m^2}V_{eff}\right)^2+\sin^2(2\theta_0) }
\label{mix}
\end{equation}
This phenomena is basically the  neutrino oscillation while propagating through a thermal heat bath filled with $\chi$ particles. For general scalar and fermion background
calculation, see Ref. \cite{Nieves:2018vxl}.

Following the sterile neutrino self-energy at one-loop in a thermal bath is given by:
\begin{align}
  \Sigma(k) = (m - a\slashed{k} - b\slashed{u}) \,.
\end{align}
Here, $m$ is the sterile
neutrino mass, $p$ is its 4-momentum and $u$ is the 4-momentum of the heat bath and is taken to be $u = (1, 0, 0, 0)$ in its rest frame.
The energy dispersion relation in the medium becomes:
\begin{align}
k^0=|\vec{k}|+\frac{m^2}{2|\vec{k}|}-b\,
\end{align}
in the UV regime, which gives us:
\begin{align}
  V_\text{eff} \equiv -b \,.
  \label{eq:Veff}
\end{align}
The coefficient $b$ can then be obtained according to the relation in \cite{Quimbay:1995jn}:
\begin{equation}
  b = \frac{1}{2 \vec{k}^2} \big[ [ (k^0)^2 - \vec{k}^2 ] \tr\,\slashed{u} \Sigma(k) 
                                 - k^0 \tr\,\slashed{k} \Sigma(k) \big] \,.
\end{equation}
To evaluate b one needs $\Sigma(k)$ which in leading order receives the thermal corrections from the bubble and the tadpole diagrams:

The leading thermal contributions to the bubble diagram are
\begin{equation}
  g ^2 _s \int\!\!\frac{d^4 p}{(2\pi)^4} \gamma^5 (\slashed{k} + \slashed{p}) \gamma^5 \bigg[ \frac{i \Gamma_f(k+p)}{p^2 + m_{\chi} ^2} - \frac{i \Gamma_b(p)}{(k+p)^2 - m_{\nu_s} ^2} \bigg] \,.
\end{equation}
$\Gamma_f(p)$ the thermal parts of the fermionic propagators:
\begin{align}
  \Gamma_f(p) &= 2\pi \delta(p^2 - m^2 _{\nu_s}) \eta_f(p) \,, \\
  \Gamma_b(p) &= 2\pi \delta(p^2 - m^2 _{\chi}) \eta_b(p) \,,
\end{align}
with corresponding $\eta_f(p)$ and $\eta_f(b)$ are the Fermi-Dirac and Bose-Einstein distribution occupation numbers respectively.

Using \cite{Enqvist:1990ad,Weldon:1982bn} first delta function integrals are done to do p$^0$ part. Then the 3-momenta p-integral is shifted to 
spherical co-ordinates which will reduce to the standard $\vec{p}$ integral which is to be performed numerically.

We give some analytic estimates for the integral for low temperature limit: 

\begin{equation}
V_{\rm eff} ^{\rm bubble}=-\frac{7 \pi^2 g ^2 _{s} E T_{\chi}^4}{180 m_{\chi} ^4} ( T_{\chi},E<< m_{\chi})
\end{equation}
\begin{equation}
V_{\rm eff} ^{\rm bubble}=\frac{ g ^2 _{s}T_{\chi}^2}{32E} (T_{\chi},E>>m_{\chi})
\end{equation}
And similarly, for tadpole diagrams,

\begin{align}
V_{\rm eff}^{\rm tadpole}\simeq\frac{ g^2 _s}{8 m_{\chi}^2}(n_{f}-n_{\bar{f}})\,,
\end{align}

The origin of the Eqn. \eqref{mix} comes from the secret "pseudoscalar" interaction which introduces a matter potential causing MSW-like effect for sterile neutrinos
of the form~\cite{Babu:1991at,Enqvist:1992}:
\begin{equation}\label{eq:Vs}
V_s(p_s) = \frac{g_s^2}{8 \pi^2 p_s} \int p dp \left(f_\phi + f_s \right),
\end{equation}
where $f_\phi$ is the Bose-Einstein distribution for the pseudoscalar and $f_s$ is the distribution for the sterile neutrinos.
The potential V$_s(p_s)$ is basically the thermal contribution of the background field in the form of bubble diagrams; an order-of-magnitude estimate of it goes as:
\begin{equation}
V_s \sim 10^{-1} \, g_s^2 T.
\end{equation}
considering a common temperature T for the all the species. This is the \textit{central idea} behind the phenomenology that this matter-effect would induce a mixing angle
different from that of the standard $\nu_s \rightarrow \nu_a$ (2-flavor approximation) and stop it from thermalizing with the SM. This can be alternatively looked upon as a minute shift
in the effective mass-difference between the neutrino states.

For a 2-neutrino framework, the thermalization process can be treated easily by the density matrix formalism leading to solving the Quantum Kinetic Equation (QKE) in
equilibrium:
\begin{equation}
\rho = \frac{1}{2}f_0
\begin{pmatrix}
P_a & P_x-iP_y\\
P_x+iP_y & P_s
\end{pmatrix},
\end{equation}
where $f_0$ is the Fermi-Dirac distribution function. 
The QKEs are now
\begin{align*}
\dot P_a &= V_x P_y + \Gamma_a \left[ 2 - P_a\right],\\
\dot P_s &= -V_x P_y + \Gamma_s \left[2 \frac{f_{\textrm{eq},s}(T_{\nu_s},\mu_{\nu_s})}{f_0} - P_s\right],\\
\dot P_x &= -V_z P_y - D P_x,\\
\dot P_y &= V_z P_x - \frac{1}{2} V_x (P_a-P_s) - D P_y .
\end{align*}
and the potentials are:
\begin{align*}
V_x &= \frac{\delta m_{\nu_s}^2}{2p} \sin 2\theta_s,\\
V_z &= -\frac{\delta m_{\nu_s}^2}{2p} \cos 2\theta_s - \frac{14\pi^2}{45\sqrt{2}} p \frac{G_F}{M_Z^2} T^4 n_{\nu_s} + V_s ,
\end{align*}
where $p$ is the momentum, $G_F$ is the Fermi coupling constant, $M_Z$ is the mass of the Z boson, and $n_{\nu_s} = \int f_s d^3p/(2\pi)^3$ is the number density of 
sterile neutrinos. The range of the values of coupling $g_s$ for which $\triangle N_{\rm eff}$ varies from 1 to 0 is $g_s\sim 10^{-6}$ to $10^{-5}$ \cite{Archidiacono:2014nda}.


\end{document}